\documentclass[reprint, amsmath, amssymb, aps]{revtex4-2}

\usepackage{graphicx}% Include figure files
\usepackage{dcolumn}% Align table columns on decimal point
\usepackage{bm}% bold math
\usepackage{subfigure}

\begin{document}

\preprint{APS/123-QED}

\title{Eigenvalue Preferential Attachment Networks\\ A Dandelion Structure}% Force line breaks with \\
%\thanks{A footnote to the article title}%

\author{V. Adami}
	\affiliation{Department of Physics, University of Mohaghegh Ardabili, P.O. Box 179, Ardabil, Iran}
	
	\author{Z. Ebadi}
	\affiliation{Department of Physics, University of Mohaghegh Ardabili, P.O. Box 179, Ardabil, Iran}

 \author{M. Nattagh-Najafi}
	\affiliation{Department of Physics, University of Mohaghegh Ardabili, P.O. Box 179, Ardabil, Iran}

\date{\today}% It is always \today, today,
             %  but any date may be explicitly specified

\begin{abstract}
In this paper we introduce a new type of preferential attachment network, the growth of which is based on the eigenvalue centrality. In this network, the agents attach most probably to the nodes with larger eigenvalue centrality which represents that the agent has stronger connections. A new network is presented, namely a dandelion network, which shares some properties of star-like structure and also a hierarchical network. We show that this network, having hub-and-spoke topology is not generally scale free, and shows essential differences with respect to the Barab{\'a}si-Albert preferential attachment model. Most importantly, there is a super hub agent in the system (identified by a pronounced peak in the spectrum), and the other agents are classified in terms of the distance to this super-hub. We explore a plenty of statistical centralities like the nodes degree, the betweenness and the eigenvalue centrality, along with various measures of structure like the community and hierarchical structures, and the clustering coefficient. Global measures like the shortest path statistics and the self-similarity are also examined. 
\end{abstract}

%\keywords{Suggested keywords}%Use showkeys class option if keyword
                              %display desired
\maketitle

\section{Introduction}
Understanding the intricate characteristics exhibited by contemporary real networks constitutes the primary focus of Complex Networks research. These distinctive attributes, such as the clustering coefficient and betweenness, defy explanation through the conventional Erd{\H{o}}s-R{\'e}nyi random graph model ~\cite{erdHos1960evolution} and regular lattice structures. In response to this challenge, Watts-Strogatz's small-world model~\cite{watts1998collective} and Barab{\'a}si-Albert's (BA) scale-free (SF) model ~\cite{barabasi1999emergence} emerged as efforts to elucidate the non-trivial properties observed in diverse networks such as computer networks, the internet, social networks, biological networks, and brain networks ~\cite{dorogovtsev2003evolution, pastor2004evolution, albert1999diameter,wasserman1994social,borgatti2009network,salwinski2004database,farahani2019application}. \\
Among these models, BA model has been put in an intense focus as the first microscopic model for SF real networks with a plenty of behaviors like power-law and scaling behaviors, which identifies an important universality class in complex networks. The mechanism of \textit{preferential attachment} in this model is used to describe the process of agent attraction within various networks ~\cite{newman2001clustering, jeong2003measuring, eisenberg2003preferential, capocci2006preferential, vazquez2003growing, kunegis2013preferential, de2007preferential, zadorozhnyi2015growing, zhou2007modelling}. In growing complex networks, the preferential attachment mechanism refers to the preference of the newly added agents to be attached to an agent with a higher \textit{centrality}, which is chosen to be the \textit{degree centrality} in BA model. The likelihood that a newly added node links to a node $i$ with degree $k_i$ is proportional to a power of its degree according to ~\cite{kunegis2013preferential, de2007preferential, zadorozhnyi2015growing, zhou2007modelling}
\begin{equation}
    \Pi^{(\text{BA})}(i)= \frac{k_i^{\zeta}}{\sum_j k_j^{\zeta}},
    \label{Eq:bapref}
\end{equation} 
where $\zeta$ is a nonlinearity exponent, which is one for the BA model with power law degree distribution~\cite{newman2001clustering, jeong2003measuring, eisenberg2003preferential, capocci2006preferential, vazquez2003growing}.

The preferential attachment algorithm, which might change from system to system, has found many applications in real networks like protein network evolution~\cite{eisenberg2003preferential}, online networks~\cite{kunegis2013preferential}, internet infrastructure~\cite{vinciguerra2010geography}, internet encyclopedia Wikipedia~\cite{capocci2006preferential}, see~\cite{barabasi2003scale,jeong2003measuring} for a good reference. To explain various real networks, a lot of variants were proposed, each of which modeling the attachment algorithm. For example, in a model termed Bianconi-Barab{\'a}si Network, the stochastisity comes to play via a fitness factor, or attractivity~\cite{bianconi2001competition}. Homophillic model~\cite{kurka2015online}, and Euclidean Distance Model~\cite{soares2005preferential} (resulting to $q$-Gaussian distribution for the degree) are the other examples. The Fitness Model and Homophilic Model with euclidean distance~\cite{nunes2017role} are mixed other models, see~\cite{piva2021networks} for more details. Re-wiring the connections is another strategy to modify the BA universality class~\cite{bertotti2019configuration}. In some growing complex networks, particularly those composed of traders (trade complex networks)~\cite{wang2019evolution,hou2018structure,bhattacharya2008international} one may expect that the preferential attachment scheme should be very effective~\cite{liu2022preferential,chen2018global}. The same is expected in Socio-political complex networks~\cite{jalili2013social}. In such networks, the addition of a new member to the system may be driven by their pursuit of more influential individuals capable of connecting it to a center of power for expediting their objectives. The sought-after centrality of the candidate does not exclusively correspond to the highest degree within the network, but rather, it can manifest through various forms of centrality. One such example is the level of communication the person maintains with effective individuals. This particular centrality can be quantitatively captured through the concept of \textit{eigenvalue centrality}, defined as the summation of the centralities of the neighbors of a given node~\cite{estrada2005subgraph}. A high eigenvalue centrality does not necessarily translate to a high degree centrality, but it means that the node under focus has \textit{very good connections}. 
In this study, we present a growth model in which preferential attachment is a linear function of the nodes' eigenvector centrality (rather than their degree centrality). Our model, which we call as a dandelion network, demonstrates similarities with a \textit{winner-takes-all} scenario which is identified by the fact that one node catches a significant proportion of all links. Such networks have a \textit{hub-and-spoke} structure, which are often identified in air transportation networks, cargo delivery networks, telecommunication networks, and healthcare organization structures ~\cite{toh1985impact, aykin1995networking, zapfel2002planning, klincewicz1998hub, elrod2017hub}, to mention a few.\\

The paper organizes as follows: In the next section we introduce our model. Section~\ref{SEC:Results} is devoted to the results that our model leads to. It contains degree and eigenvalue statistics, central point dominance, degree dynamics, finite size scaling, shortest path and closeness statistics, clustering coefficient, assortativity and community structure, and self-similarity. We close the paper with a discussion section.

\section{Model}\label{SEC:model}

In this section we describe our preferential attachment network model, in which the nodes with greater eigenvector centralities are more likely to be picked as target nodes by newly added nodes at each time step. We define an external integer $m$ in our model which is the number of links added upon adding one node to the system. The attachment probability in our model is defined via
\begin{equation}
    \Pi(i)= \frac{v_i^{\text{max}}}{\sum_j v_j^{\text{max}}},
    \label{Eq:ourpref}
\end{equation}
which should be compared with Eq.~\ref{Eq:bapref}. In this equation $v_i^{\text{max}}$ as the eigenvector centrality of node $i$, defined as the eigenvector of the adjacency matrix corresponding to maximum eigenvalue. More precisely, if we represent $A$ as the adjacency matrix of a unweighted, undirected network with the entries $a_{ij}=1$ if $i$ and $j$ are connected, and zero otherwise, then
\begin{equation}
(A-I\lambda) \textbf{v}=0.
\label{Eq:ev_eqt}
\end{equation}
If we show the maximum eigenvalue as $\lambda_{\text{max}}>0$ (the positivity is guaranteed by the Perron-Frobenius theorem), then this relation can be written as
\begin{equation}
	v_i^{\text{max}}=\frac{1}{\lambda_{\text{max}}}\sum_{j\in K_i} v_j^{\text{max}}.
	\label{Eq:v_i}
\end{equation}
where $K_i$ is the set of neighbors of $i$, i.e. $a_{ij}=1$. For simplicity, from now on we show the eigenvectors corresponding to the maximum eigenvalues as $v_i\equiv v_i^{\text{max}}$. This relation explains why the eigenvector centrality of node $i$ is about the strength of connections that it has. \\
The model is defined as follows: We start with a fully connected network with $m+1$ nodes, and add nodes one by one to the system. In each step, the newly added node to the system establishes $m$ links to \textit{different} nodes according the probability given in Eq.~\ref{Eq:ourpref}. Note that no node can receive more than one link from a given node, so that the graph is a simple graph rather than a multi-graph. Therefore, in each times step $t$, we find the eigenvalues and eigenvectors of the $t\times t$ adjacency matrix $A$. The centrality of the nodes that we concern are the one corresponding to the maximum eigenvalue. It is important to have in mind that the eigenvector centrality of all nodes are updated upon adding a new node even if no connection is established to it. Note that to have a unique centarlity for each site, we need normalization in each time step as follows
\begin{equation}
	\sum^N_i v^2_i =1.
    \label{Eq:norm}
\end{equation}
In connected networks with the number of nodes greater than two the maximum possible value of $v_i$ is $\frac{\sqrt{2}}{2}$, which corresponds to the highest central node. The system size $N$ is defined as the final time of the simulations ($N\equiv t_{\text{final}}$).\\
 
\section{The network structure}~\label{SEC:Results}
\begin{figure}[t]
    \includegraphics[width=7cm]{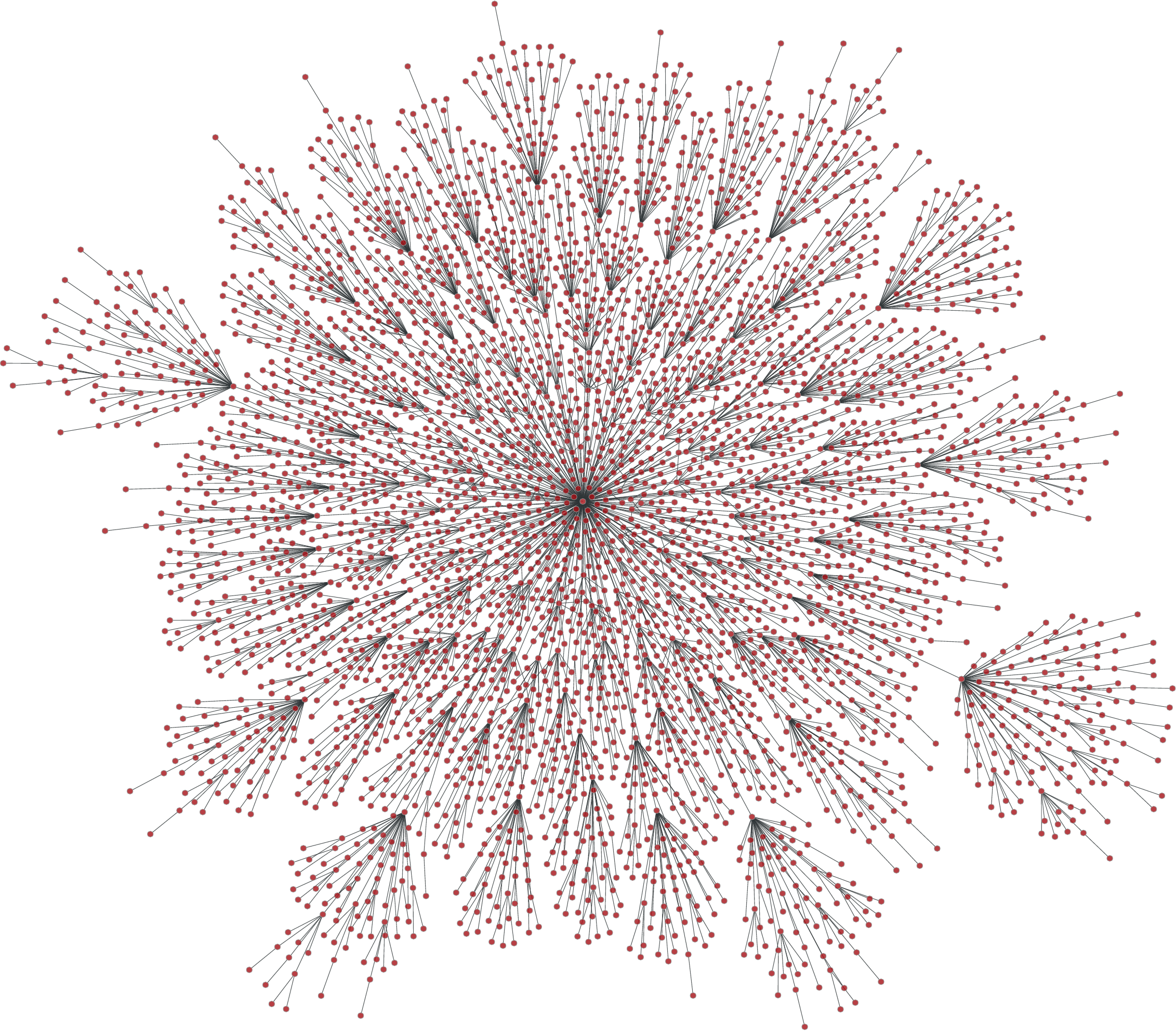}
    \caption{\label{fig:dandelion}A dandelion network with $N=4000$ nodes and $m=1$.}
\end{figure}
This section is devoted to the simulations results. For the simulations, we have used Python libraries Graph-Tool~\cite{peixoto_graph-tool_2014} and Networkx~\cite{SciPyProceedings_11}. In the following subsections we analyze the network measure one by one, and compare them with the same quantities in the BA is available.\\
Figure~\ref{fig:dandelion} depicts a sample network for $m=1$, representing a dandelion structure with the so-called hub-and-spoke topology. There is a super-hub at the network's core with the highest eigenvector centrality value of $\approx \frac{\sqrt{2}}{2}$. In the remainder of this paper, we introduce this structure and statistically investigate its properties.

\subsection{Degree and eigenvector centralities}
\begin{figure*}[t]
  \centering
  \subfigure[\label{fig:16000_ensemble}]{\includegraphics[scale=0.86]{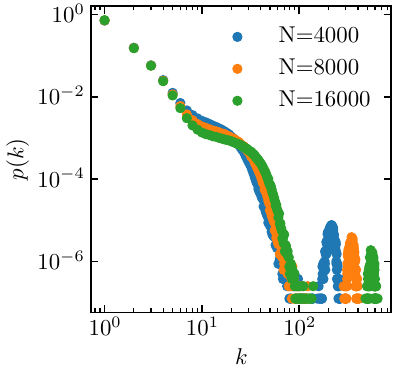}}\,
  \subfigure[\label{fig:ev_distribution_1}]{\includegraphics[scale=0.86]{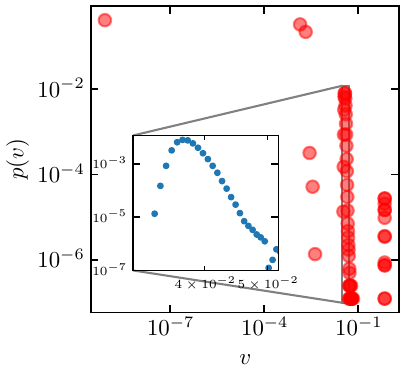}}\,
  \subfigure[\label{fig:e_k_1}]{\includegraphics[scale=0.86]{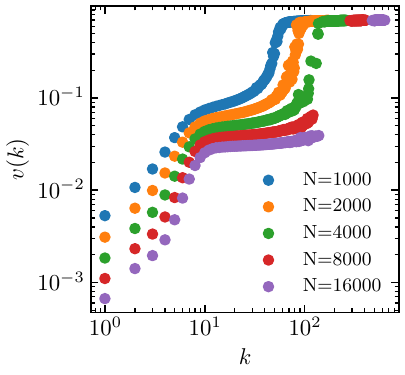}}
      \caption{(a) Degree distribution of networks with sizes of 4000, 8000, and 16000 realized from 2000, 1000, and 500 samples, respectively, assuming $m=1$ for all of them. Peaks represent the degree distribution of each sample's super-hubs, whose degrees deviate significantly from those of other nodes. As the size of the networks grows, the pace at which super-hubs' degrees increase is substantially faster than the rate at which other nodes' degrees increase. (b) Eigenvector centrality distribution of networks with $N=16000$ and $m=1$ parameters. This distribution can be divided into four sections. The inset is a zoomed-in portion of the second part from the right that reveals a peak for a certain value of the eigenvector centralities. (c) the eigenvector centrality feature of our model as a function of node degree for different network sizes of 1000, 2000, 4000, 8000, and 16000, all with the same parameter of $m=1$.}
    \label{fig:degree_eigenvector}
\end{figure*}

The node degree $k$ ($\#$ links that end up in a given node) and the eigenvector $v$, and the corresponding distribution functions ($p(k)$ and $p(v)$ respectively) are analyzed in this section. Figure~\ref{fig:16000_ensemble} shows the degree distribution of three different network sizes. First observe that these functions do not show power-law behavior, they show two separate peaks, i.e. our network is not SF. Additionally they comprise of two parts which move away as $N$ increases. The discontinuities and the observed gap are caused by the degree difference between the super-hubs (the right part of the plot) and the remaining nodes. This super-hub is the central point in Fig.~\ref{fig:dandelion}, and the smooth (left) peak that is observed in intermediate $k$ values is due to the accumulation of nodes that are directly connected to the super-hub. As the network grows in size, the degree difference between the super-hubs and the rest of the nodes becomes more pronounced. This implies that super-hubs attract new entering nodes at a faster pace than regular nodes.

The distribution function of the eigenvector centrality is shown in Fig.~\ref{fig:ev_distribution_1}, according to which the separated parts are observed: the most important nodes or super-hubs have the highest eigenvector centrality value of almost $\simeq\frac{\sqrt{2}}{2}$.

\begin{figure}[b]
  \centering
  \subfigure[\label{fig:BA_ev_dist}]{\includegraphics[scale=0.6]{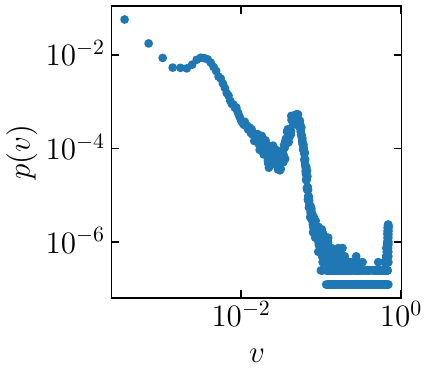}}\,
  \subfigure[\label{fig:BA_v_k}]{\includegraphics[scale=0.6]{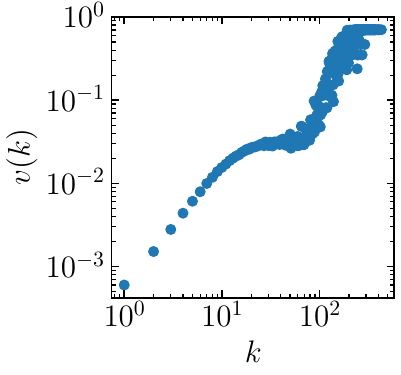}}
      \caption{(a) Eigenvector centrality distribution of the BA network with the parameters $N=8000$ and $m=1$. (b) the eigenvector centrality feature as a function of the nodes' degree of the BA model for the network size of 8000 given $m=1$. None of these distributions show discontinuities.}
    \label{BA_dists}
\end{figure}
The first neighbors of the super-hub (called the second-level hubs) form the second peak which occur in the intermediate $k$ and $v$ values (shown in the figure \ref{fig:ev_distribution_1} as a zoomed inset). As the network grows in size, the distribution width of this group of nodes narrows. The peak in $p(v)$ is explained in terms of the plateau in the $v-k$ plot, Fig.~\ref{fig:e_k_1}, leading to an accumulation of points in the peak point. The plateau forms due to the direct connection to a super-hub node. One may use Eq.~\ref{Eq:v_i} to explain this flatness, according to which the eigenvector centrality of a second-level hub ($i$) in the case of $m=1$ is found to be 
\begin{equation}
v_{i\in \ \text{second level}} = \frac{1}{\lambda_{\text{max}}}\left[v_{\text{sh}}+\sum_{j\in \ \text{third level}}a_{ij}v_j+...\right],
\end{equation}
where $v_{\text{sh}}$ is the eigenvector centrality of the super hub, and the summation is over all nodes with distance 2 from the super hub. The number of links that are connected to the super hub (=$\#$ second level nodes) is roughly $N^{\frac{1}{2}}$, and the number of third level nodes is $\frac{1}{2}N$, so that the average degree of the second level nodes is $\frac{1}{2}N^{\frac{1}{2}}$. Given that $v_{j\in \ \text{third level}}\approx 10^{-3}$ for $N=10000$, we conclude that the second term in the bracket is of order $5\times 10^{-2}$ for $N=10000$, which is pretty smaller that the first term which is of order $\sqrt{2}/2$. We calculate the amount of the plateau in SEC.~\ref{flatness} for a star graph, and show that it coincide with our results. \\   

In Figure \ref{BA_dists} we show the same quantities for the Barab{\'a}si-Albert networks. While it turns out that the eigenvalue centrality first saturates, and then undergoes a boost to a new value (with some corresponding peaks in the distribution function), there is no discontinuity in the graph. Such discontinuities are observed e.g. in Figure \ref{fig:e_k_1}, which shows that increasing the system size leads to disparities in the degree and eigenvector centralities of the super-hubs on one hand and the remainder of the nodes on the other. Such discontinuities will appear in networks with central point dominance (CPD) values near to one (see to the section \ref{cpd}). 
\begin{figure*}[t]
  \centering
  \subfigure[\label{fig:b_k}]{\includegraphics[scale=0.89]{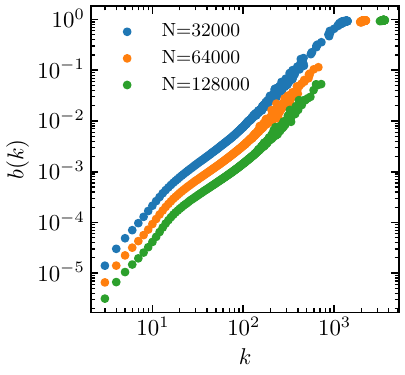}}\,
  \subfigure[\label{fig:cpd_BA}]{\includegraphics[scale=0.89]{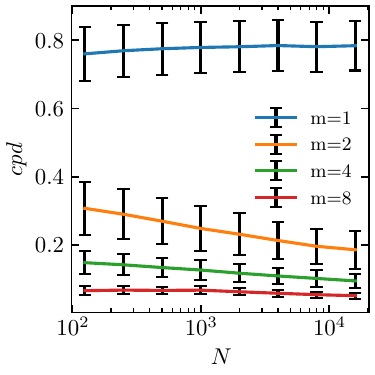}}\,
  \subfigure[\label{fig:cpd_ec}]{\includegraphics[scale=0.89]{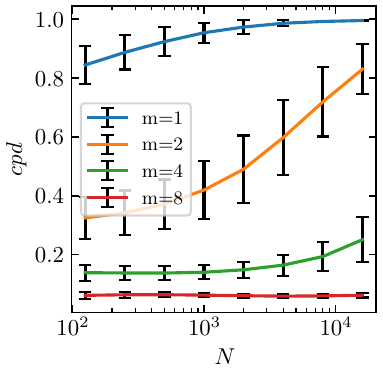}}
      \caption{(a) Betweenness centrality as a function of the degree of the nodes for three network sizes of 32000, 64000 and 128000 keeping  $m$ fixed as $m=2$ for all of them. Super-hubs have the highest betweenness centrality. There is also a meaningful difference between the betweenness centrality of the super-hubs and that of the other nodes. The black solid line has a slope of $\simeq 2.02$ and fitted to the network with the size of 128000. The global CPD feature in terms of the different system sizes for the (b) BA model and (c) our model. The BA model keeps its initial structre; i.e. its CPD value doesn't change while this isn't true for our model. By increasing the size of the system, the structure of the networks is gradually converted to a star-like structure that is the CPD value tends to one.}
    \label{fig:cpd_figs}
\end{figure*}

\subsection{\label{cpd}Central point dominance}

As a global feature, CPD is defined ~\cite{freeman1977set} as
\begin{equation}
\text{CPD}=\frac{1}{N-1}\sum_{i}(B_{\text{max}}-B_{i}),
\end{equation}
where $B_{\text{max}}$ is the network's greatest value of betweenness centrality, and $B_i$ represents the betweenness centrality of node $i$. 

This feature is used to provide a perspective on network topology in such a way that networks with a CPD value of zero (the lowest possible value) have a complete graph structure, whilst networks with a CPD value of one (the greatest possible value) have a star graph structure. The greater the CPD value of a network, the higher the number of hubs in that network. Networks with a CPD value of one have only one hub $B_{\text{max}}=B_{\text{hub}}=1$, and $B_{i\ne \text{hub}}=0$ which is a star graph. We call such a node as the super-hub. Figure \ref{fig:b_k} shows betweenness in terms of the node degree for our model. For finite $N$ values there are two distinct regions with two different slopes (exponents), while a super hub is distinguishable as a separate region which is separated from the others by a gap. While the network is not scale free (SF) in a general sense, in the mentioned two regimes we see different power-law behaviors like $b^{(i)}\propto k^{\eta^{(i)}}$ where $i=1,2$ counts the number of the intervals, and $\eta^{(1)}=1.374\pm 0.008$ and $\eta^{(2)}= 2.04\pm 0.02$ for $N=128000$ and $m=2$. This should be compared with the Barab{\'a}si-Albert model in which $b^{(\text{BA})}\propto k^{\eta_{\text{BA}}}$, where $\eta_{\text{BA}}=1.68\pm0.03$ in our calculations with the same parameters. The dependence of CPD on $N$ is shown in Fig.~\ref{fig:cpd_ec}, from which we observe how CPD approaches $1$ as $N\to \infty$ for $m=1$. This should be compared with the BA model (Fig.~\ref{fig:cpd_BA}) where CPD shows a different trend. Note that for $m>1$ the graphs tend to increase and approach $1$ for larger $N$ values, but the faith of the system needs runs for larger $N$ values which was not accessible for us. This result reveals that the topology of the networks varies by time, exhibits asymptotically a structure that, concerning CPD characteristics is similar to a star graph (at least for $m=1$), both of which have a super-hub the statistics of which is different from the rest (represented often by a gap in the spectrum). This super-hub node can also be identified by the fact that it is closest to all other nodes in the network (see section \ref{SEC:diameter_closeness}). It is worth noting that the similarity to star-like graphs is restricted to CPD and only for very large times. Although the dandelion network structure is topologically different from hub-and-spoke networks, asymptotically it exhibits properties like them where super-hub plays the role of the central agent~\cite{elhedhli2005hub,de2009multiple,bryan1999hub,o1998geographer,brueckner1992fare}.  

\subsection{Assortativity}

\begin{figure}[b]
    \includegraphics[]{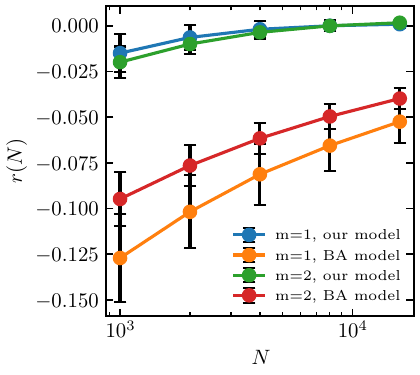}
    \caption{\label{fig:assor}Assortativity as a function of the system size.}
\end{figure}

The tendency of nodes to connect to other nodes of the network with similar degrees is called assortativity. In contrast disassortatiivty refers to a state in which nodes with low degrees prefer to link to high degree nodes, and vice verca. There can also be intermediate states in which there are no correlations between low degree and high degree nodes. Pearson correlation coefficient of the degrees~\cite{newman2002assortative} is used in order to identify this quantity. The assortativity is defined as

\begin{widetext}
\begin{equation}
	r=\frac{M^{-1}\sum_{j>i}k_ik_ja_{ij}-[M^{-1}\sum_{j>i}(1/2)(k_i+k_j)a_{ij}]^2}{M^{-1}\sum_{j>i}(1/2)(k^2_i+k^2_j)a_{ij}-[M^{-1}\sum_{j>i}(1/2)(k_i+k_j)a_{ij}]^2},
\end{equation}
\end{widetext}
with $M$ as the total number of the edges in the network. Assortativity varies in the range of $r=[-1,1]$. The network is assortative if $r>0$ and disassortative if $r<0$; there is no correlation between degrees of nodes when $r=0$. It is not surprising that the hub-and-spoke networks are dissortative~\cite{newman2003mixing}.

We have calculated this feature for both the BA and our model for two values of $m=1$ and 2. Figure \ref{fig:assor} uncovers the fact that the assortativity is negative for finite $N$ values for both models (BA is more disassortative than ours meaning that the low $k$ nodes are more demand to connect to higher $k$ nodes and vice versa, (see Ref. ~\cite{shergin2021assortativity} for more details). The assortativity tend asymptotically ($N\to \infty$) to zero for these two models in large enough networks regardless of the value of $m$. This means that the node degrees are uncorrelated in this limit.

\subsection{Community and hierarchical structure}\label{SEC:community}

\begin{figure*}[t]
  \centering
  \subfigure[]{\includegraphics[scale=0.28]{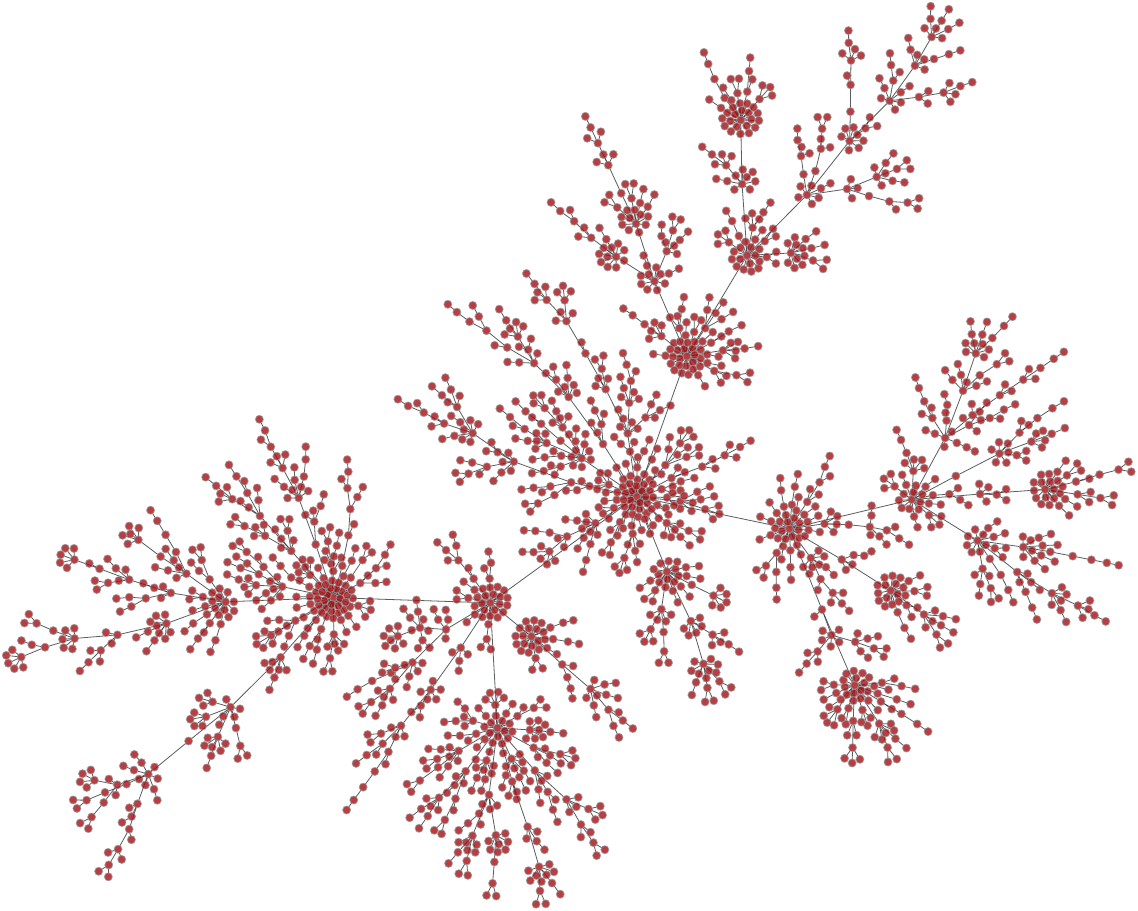}}\quad
  \subfigure[]{\includegraphics[scale=0.28]{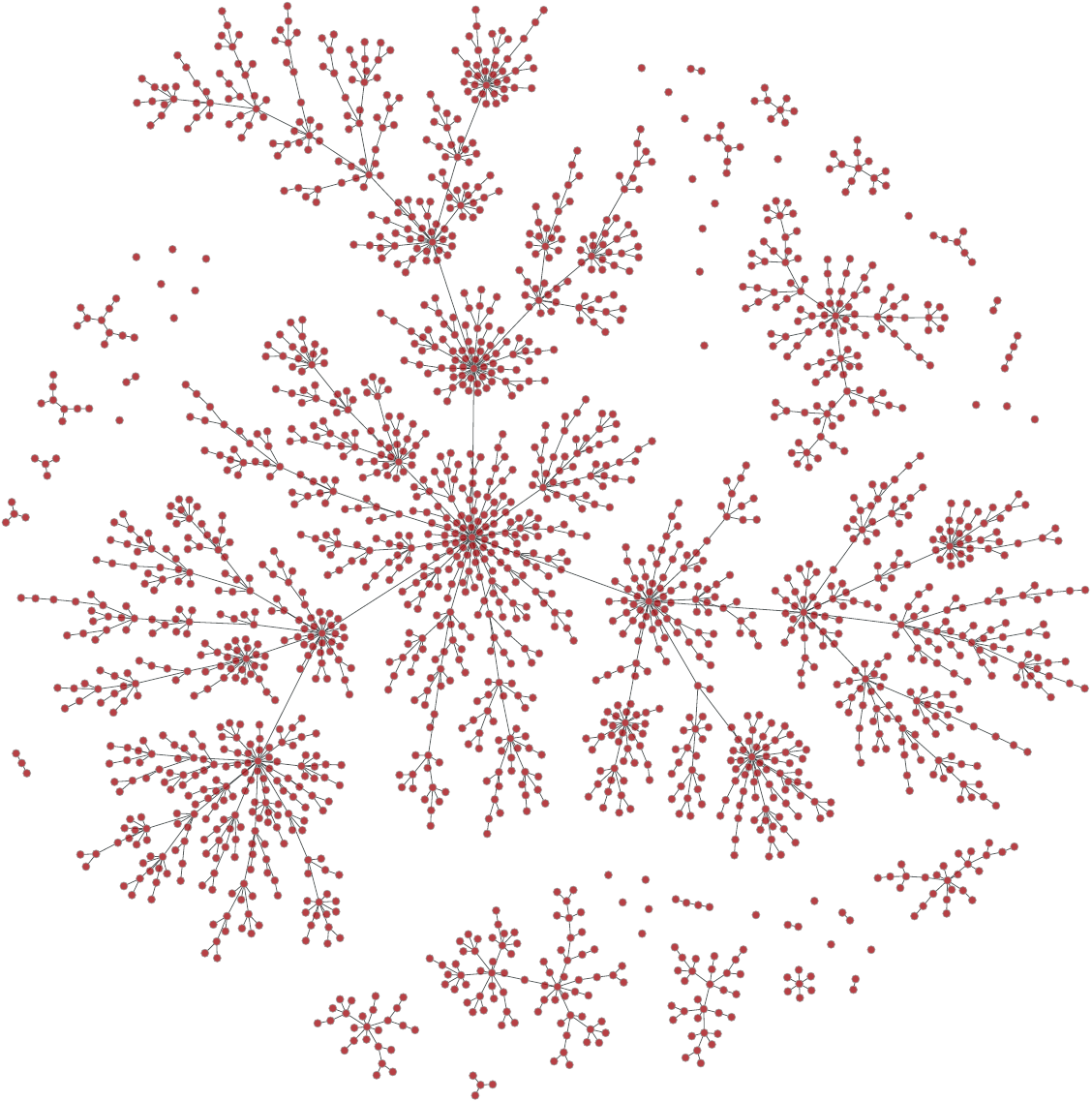}}\quad
  \subfigure[]{\includegraphics[scale=0.28]{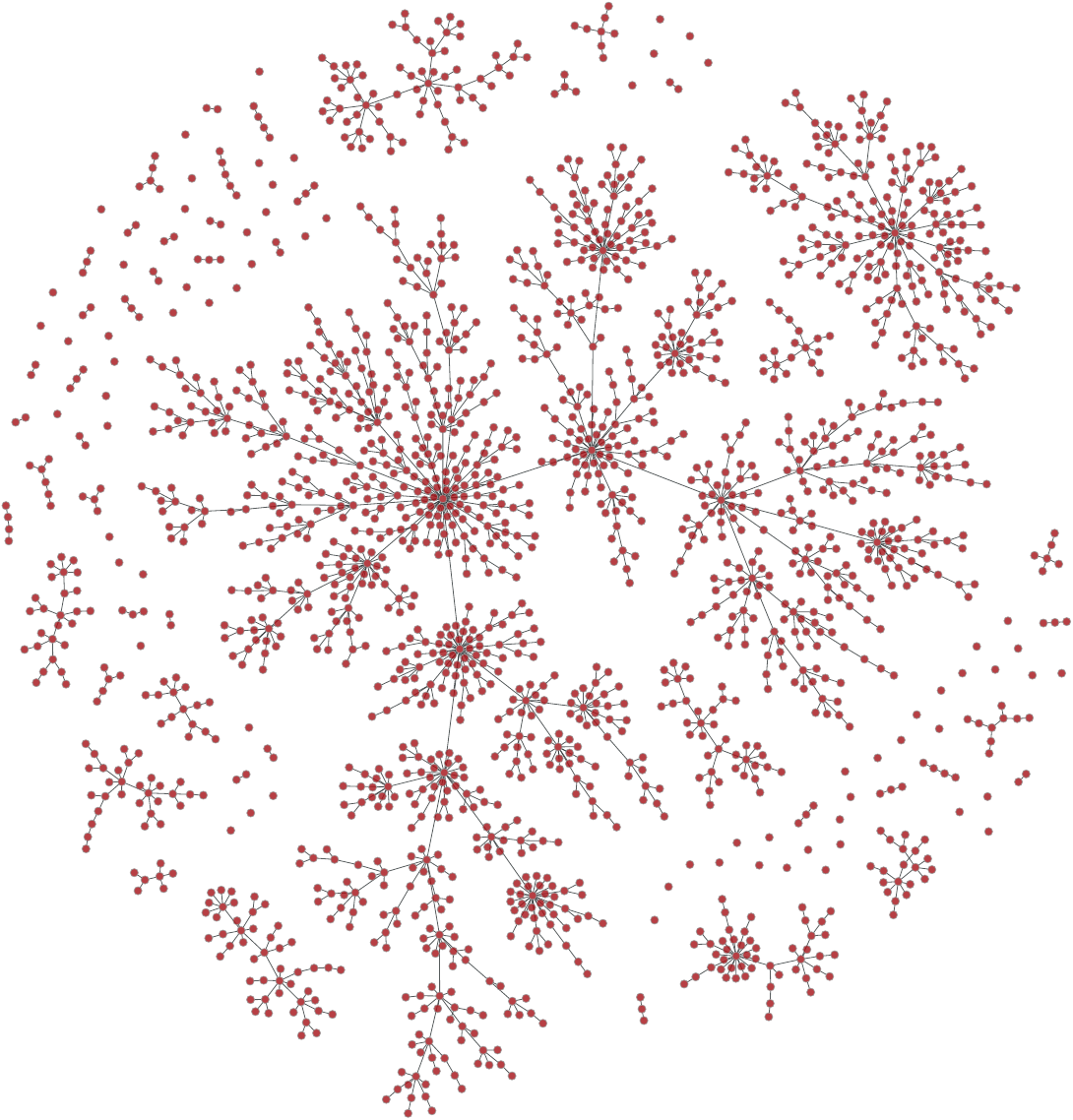}}
      \caption{A visualization of (a) level-0, (b) level-1, (c) level-2 trimmed BA model with $N=2000$ nodes and $m=1$. Note that at level $1$ the main hub of the system and its links have been removed and at level 2 the communities have been removed. See the text for more details.}
    \label{comm1}
\end{figure*}

In this analysis, we examine community and the hierarchical structures of both the BA model and our model. This comparison allows us to gain insights into the community structure perspectives of dandelion and BA networks.

A qualitative definition defines communities as network sub-graphs in which the number of links inside them exceeds the number of edges connecting them. Taking the preceding requirements into account, the \textit{modularity} parameter, $Q$, was proposed by Newman and Girvan ~\cite{newman2004finding} as follows
\begin{equation}
    Q = \sum_{i=1}^N \left[e_{ii}-\left(\sum_j e_{ij}\right)^2\right].
\end{equation}
In this equation, if the index $i$ counts the number of communities found in the network, $e_{ii}$ is the fraction of links within the community $i$, $e_{ij}$ is the fraction of links connecting communities $i$ and $j$. Our model does not show community structure, i.e. there is one community with $Q=1$ for all $m$ values.

The hierarchical structure shows a clearer structure in our model. It refers to how nodes are organized into groups or clusters with varying levels of granularity or resolution. A network with a hierarchical structure can be broken down into smaller subnetworks, and these subnetworks can further be subdivided into even smaller ones. This hierarchical arrangement can provide valuable information about the modularity, functionality, and evolution of the network~\cite{barabasi2003scale}. 
For the special case of $m=1$, the method by which we obtain the hierarchy and the corresponding communities is to remove communities from the network in a descending arrangement, i.e. we first remove super-hub ($n=1$), which results to $n_1$ separate connected graphs. Then we proceed by removing the nodes with distance $1$ from the super hub ($n=2$) resulting to $n_2$ separate connected graphs, etc. The number of communities in the $m$th level ($n_m$) shows the number of communities in that level of hierarchy. At each level, the degree of the hub determines the number of communities in that branch. We continue this process until there are no communities left in the network. A single node is not considered as a community, i.e.
we define subgraphs to be communities with at least one link between their nodes, so that the minimum size of the communities is two. \\

\begin{figure*}[t]
  \centering
  \subfigure[]{\includegraphics[scale=0.28]{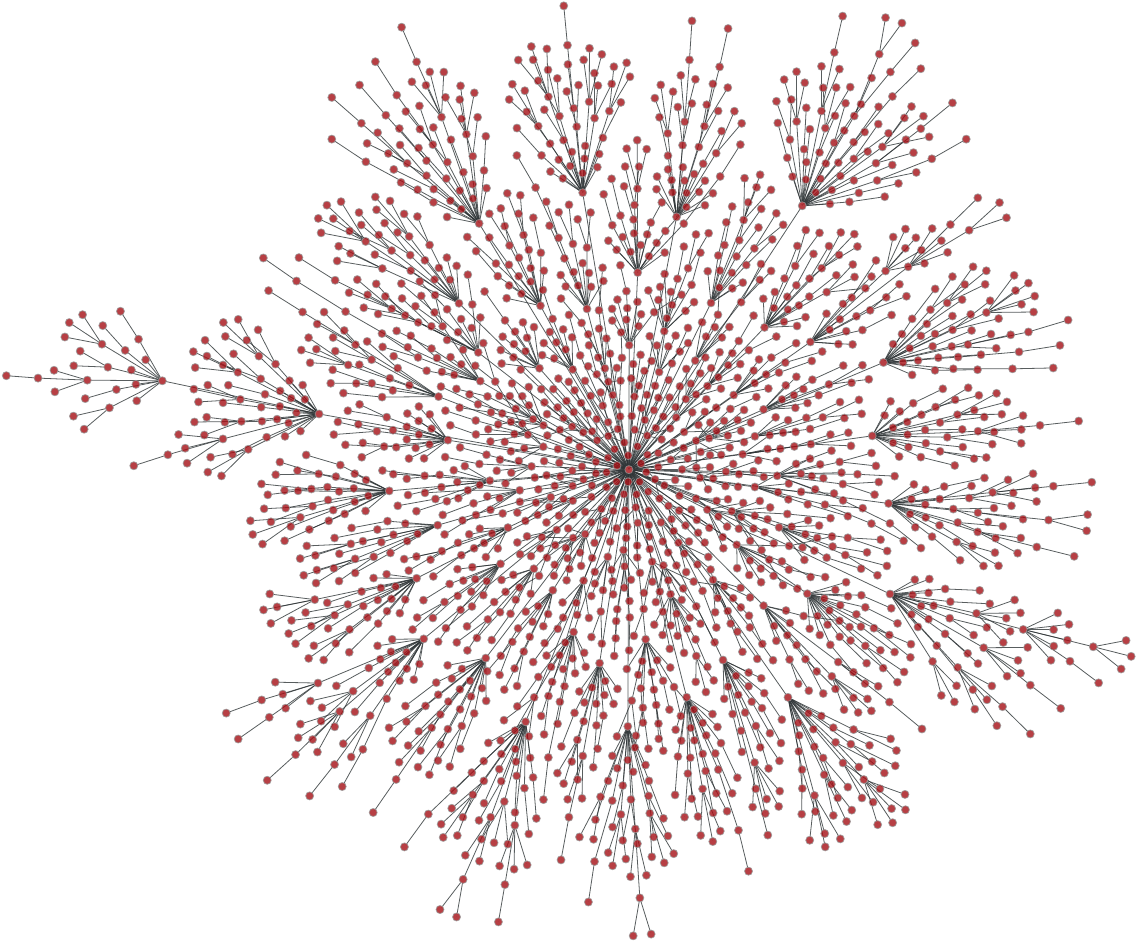}}\quad
  \subfigure[]{\includegraphics[scale=0.28]{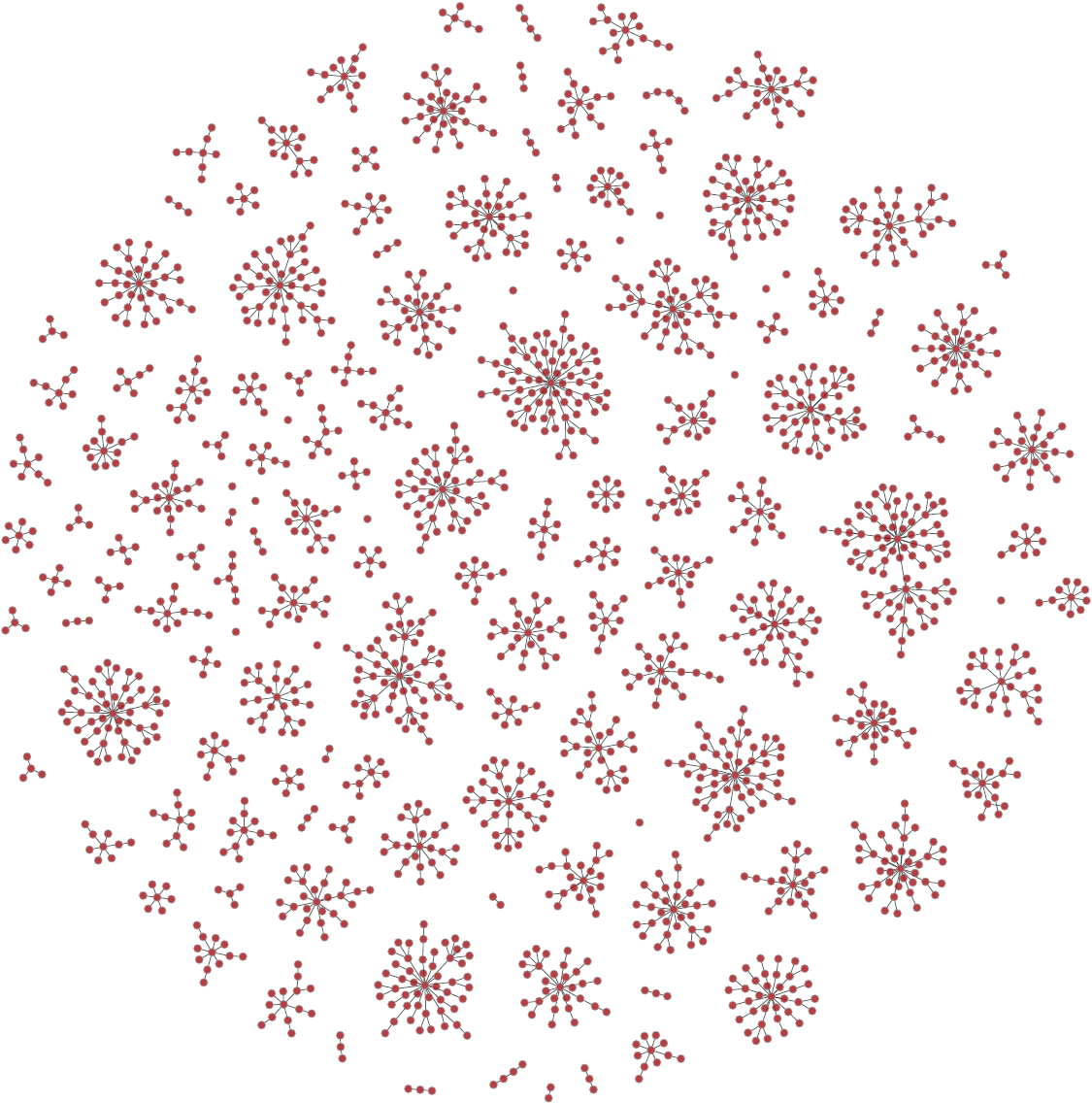}}\quad
  \subfigure[]{\includegraphics[scale=0.28]{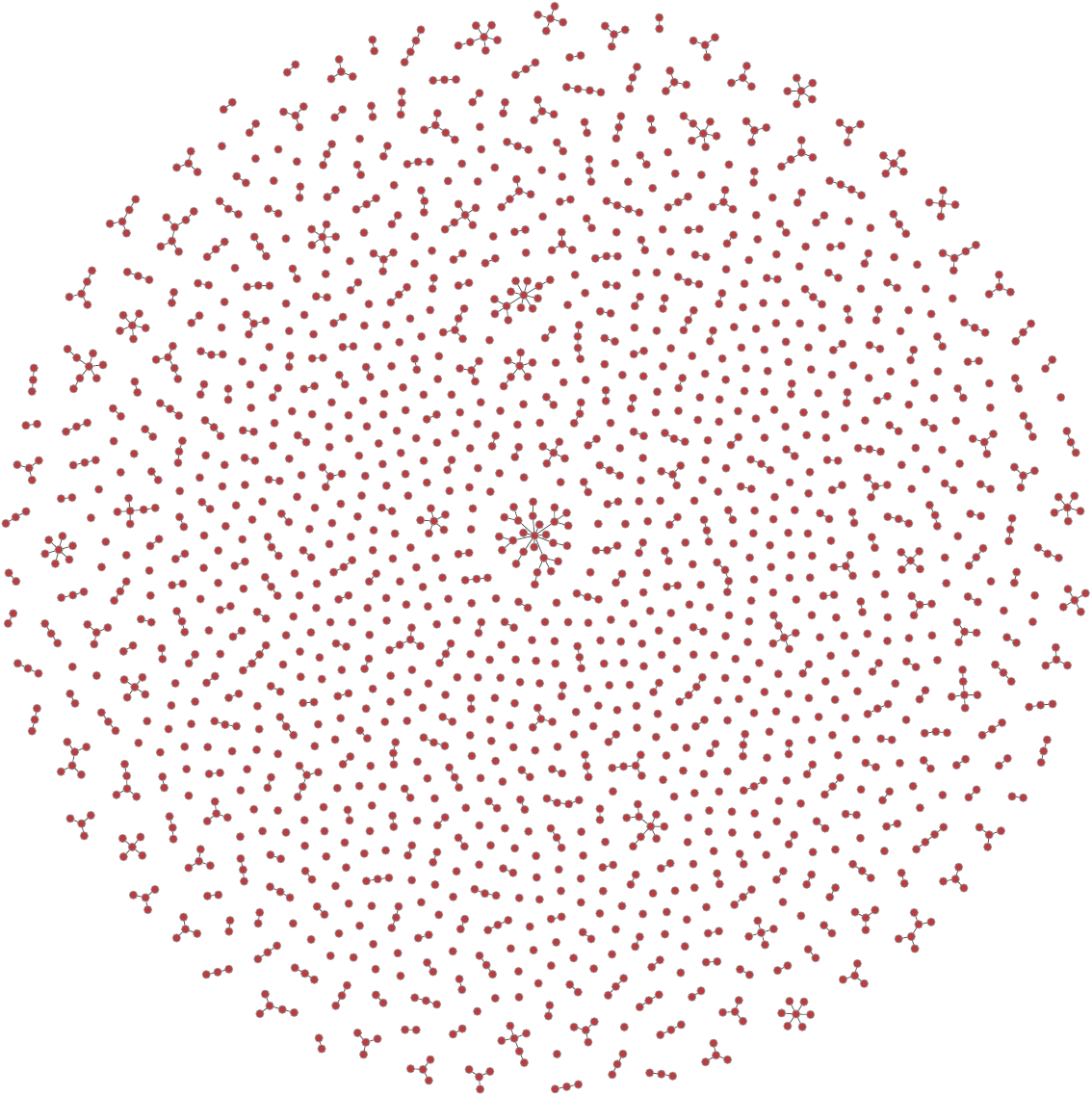}}
      \caption{The same as Fig.~\ref{comm1} for (a) level-0, (b) level-1, and (c) level-2 trimmed dandelion structure.}
    \label{comm2}
\end{figure*}

\begin{figure}[h]
  \centering
  \subfigure[]{\includegraphics[scale=0.64]{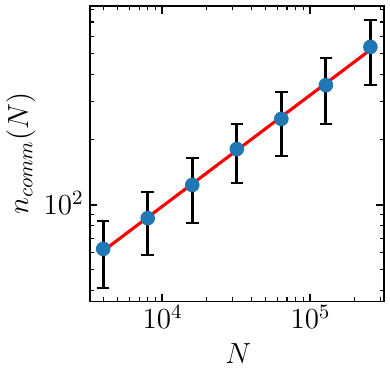}}\,
  \subfigure[]{\includegraphics[scale=0.64]{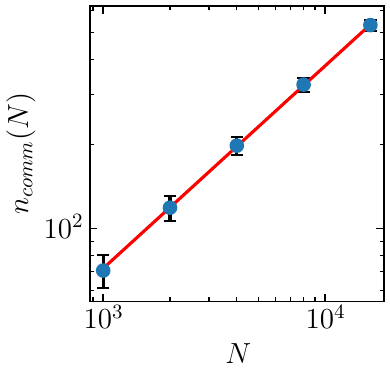}}
    \caption{Number of the communities as a function of the size of the network calculated only for the first level for (a) BA model and, (b) our model. This number, as a result of fitting shown by the red line, reveals a power-law behaviour in both models with the exponents $0.51\pm 0.09$ and $0.72\pm 0.04$ respectively.}
    \label{comm3}
\end{figure}

\begin{figure}[h]
  \centering
  \subfigure[]{\includegraphics[scale=0.6]{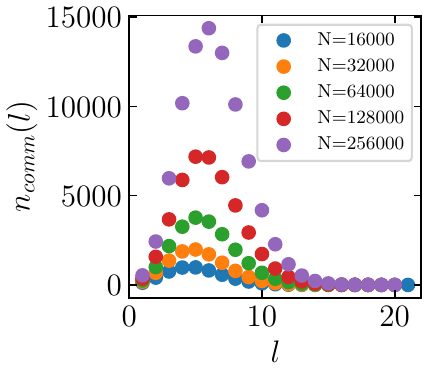}}\,
  \subfigure[]{\includegraphics[scale=0.6]{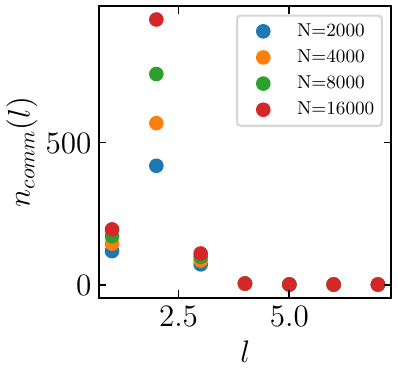}}
    \caption{Number of the hierarchical communities as a function of the different levels of obtaining communities for different network sizes; (a) BA model and, (b) our model. While this number has a maximum in level 2 for our model, it is size dependent for the BA model. Larger networks show the maximum number of the communities in higher levels.}
    \label{comm4}
\end{figure}

In the figures \ref{comm1} and \ref{comm2}, the process of removing the main hubs of system in two levels for the BA model and our model has been shown. Figure \ref{comm3} shows the number of the hierarchical communities appeared at the first level as a function of the size of the networks. This number reveals a power-law behaviour for both models
\begin{equation}
N_{\text{comm}}\propto N^{\beta},
\end{equation}
with the exponents $\beta_{\text{BA}}=0.51\pm 0.09$ and $\beta_{\text{dandelion}}=0.72\pm 0.04$. The result for the BA model is rather expected given that the node degree growth exponent of this model is $\beta=\frac{1}{2}$~\cite{barabasi1999mean}, while this exponent is expected to be level-dependent in our model.
Figure \ref{comm4} displays the number of the hierarchical communities at different levels where we have fixed the size of the network for both models. It is worth noting that the decreasing behavior (large $n$ values) is due to removing the communities with just one member. While this number has a maximum in level 2 for our model, it is size dependent for the BA model. Larger networks show the maximum number of the communities in higher levels.

\subsection{Clustering coefficient and hierarchical organization}
Clustering coefficient in the graph theory is a measurement of how closely connected nodes in a graph tend to be (for $m>1$). Most real-world networks, especially social networks, have nodes that naturally form close-knit clusters with a high density of ties. There are two versions of this measurement: local and global. The local version shows how embedded single nodes are, while the global version is intended to show how the network is clustered overall. The local clustering coefficient is defined ~\cite{watts1998collective} as
\begin{equation}
    c_i=\frac{2n_i}{k_i(k_i-1)},
\end{equation}
where $n_i$ is the number of edges between the neighbors of node $i$. The global clustering coefficient ~\cite{newman2001clustering} is defined as
\begin{equation}
    C=\frac{3N_{\triangle}}{N_3},
\end{equation}
where $N_{\triangle}$ and $N_3$ are the total number of loops of length three and the number triplets in the network, respectively. Ravasz and Barab{\'a}si ~\cite{ravasz2003hierarchical} discuss that many natural and social networks exhibit two generic properties: they are SF and have a high degree of clustering. They show that these two characteristics are the result of hierarchical structure, suggesting that small groups of nodes aggregate into increasingly large groups while maintaining a SF topology.

\begin{figure*}[t]
  \centering
  \subfigure[\label{fig:clus_degree}]{\includegraphics[scale=0.62]{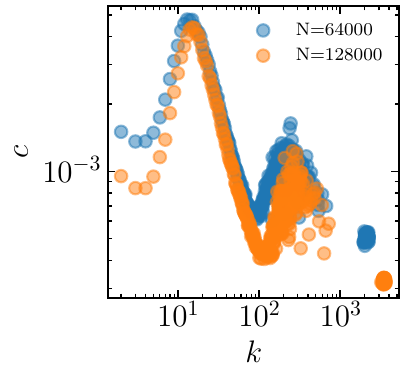}}\,
  \subfigure[\label{fig:hiararch_ccd}]{\includegraphics[scale=0.62]{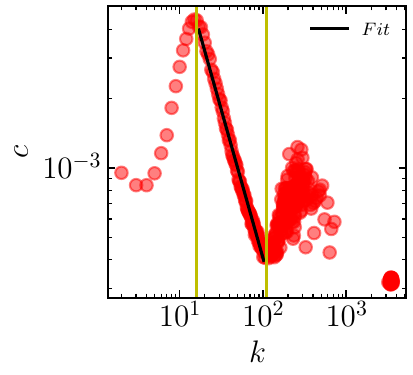}}\,
  \subfigure[\label{fig:hiararch_d}]{\includegraphics[scale=0.62]{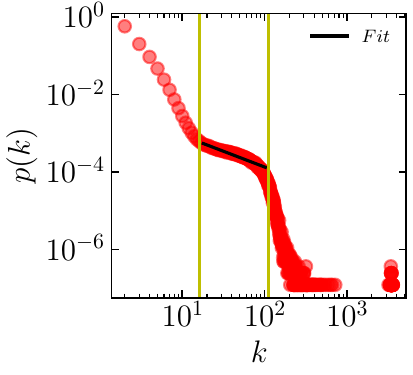}}\,
  \subfigure[\label{fig:clus_N}]{\includegraphics[scale=0.62]{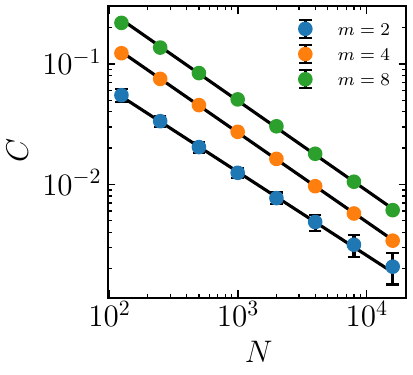}}
      \caption{(a) Local clustering coefficient as a function of the degree of nodes for the network sizes of 64000 and 128000 given $m=2$; there are some minimums and maximums. These two features are correlated in our model and reveal a power-law behaviour with constant exponents in large enough networks. Super-hubs have the lowest value of the local clustering coefficient.
      (b) We have isolated the largest network ($N=128000$) of the previous subfigure ((a)) and specified two values of its degrees by two vertical yellow lines. The local clustering coefficient reveals one maximum and one minimum in terms of the degrees identified by the left and the right yellow lines, respectively. The local clustering coefficient has a scaling power-law relation in terms of the degrees of the interval made by those yellow lines with the slope of $-1.28 \pm 0.19$.
      (c) The degree distribution of the network presented in (b). Those two specific degrees have been identified here again by two vertical yellow lines. These two special degrees specify values in terms of them, the degree distribution reveal two peaks. The degree distribution is not power-law in general but one can find the degree distribution to be scale-free within the specified interval with the slope of $0.81 \pm 0.02$.
      (d) Global clustering coefficient as a function of the size of the networks for different values of $m$; the black ($m=2$), green ($m=4$) and blue ($m=8$) lines indicate the slope of $-0.69 \pm 0.04$, $-0.722 \pm 0.004$ and $0.701 \pm 0.007$ respectively.}
    \label{fig:clus_hiararch}
\end{figure*}

Figure \ref{fig:clus_degree} depicts the result of the local clustering coefficients as a function of the nodes' degree. In contrast to the BA model in which these two features are uncorrelated, they are correlated in our dandelion structure. 
previously in the Figs.~\ref{fig:b_k},~\ref{fig:hiararch_ccd}, and ~\ref{fig:hiararch_d}, according to which
\begin{equation}
c(k)\sim k^{-\gamma_{ck}},
\end{equation}
where $\gamma_{ck}=1.3\pm 0.2$ for $10\lesssim k\lesssim 100$, and other values in the other intervals, especially it is positive for $k\lesssim 10$ and $k\gtrsim 100$. The reason of formation of these intervals is the different structures for various hierarchical communities. For example the intermediate $k$ values corresponds to the nodes which are immediately connected to the super-hub node. The dynamics and structure of these nodes are different from the other communities like the super-hub nodes, or the high $n$-level communities. Again, we observe that while the dandelion network is not a SF networks, it shows power-law behavior in distinct intervals. \\

The global clustering coefficients analysis demonstarte the following power-law behavior
\begin{equation}
C\propto N^{-\gamma_{CN}},
\end{equation}
where $\gamma_{CN}=0.69\pm 0.04$ (for $m=2$), which should be compared with the exponents of the BA ($\gamma_{CN}^{(\text{BA})}=\frac{3}{4}$~\cite{albert2002statistical}) and Erd{\H{o}}s-R{\'e}nyi ($\gamma_{CN}^{(\text{ER})}=1$~\cite{albert2002statistical, dorogovtsev2002evolution}) models.

\section{Global measurements}~\label{SEC:global}
Understanding the global properties of complex networks is crucial as they unveil the fundamental patterns, mechanisms, and functionalities that the networks represent. By global properties we mean the properties that are related to non-local observables and deal with the walks over the edges over the network. This includes path length statistics, closeness, and the re-scaling self-similarity.

\subsection{Average shortest path length and closeness}\label{SEC:diameter_closeness}
The average number of links along the shortest paths for all possible pairs of network nodes is known as average shortest-path length, which is an important measure in network topology. For undirected graphs, the average shortest path length is defined as
\begin{equation}
	l=\frac{2}{N(N-1)}\sum_{j>i}d_{ij},
	\label{avereage_distance1}
\end{equation}
where $d_{ij}$ is the shortest path length between nodes $i$ and $j$. The average shortest path length of most well-known networks, such as the BA SF network for which $l\sim \frac{\ln N}{\ln(\ln N)}$~\cite{fronczak2004average}, ER network for which $l\sim \ln N$~\cite{chung2001diameter, albert2002statistical}, and Watts-Strogatz small-world network for which $l\sim N$ if $N<N^*$ and $l\sim \ln N$ if $N>N^*$ with $N^*$ as $p$-dependant crossover size separating the small and large world regimes ~\cite{scala2001small} increases by the size of network. Indeed, the behavior of the average shortest path length is a gauge for the universality class of the complex networks~\cite{costa2007characterization}. \\

\begin{figure}[t]
  \centering
  \subfigure[\label{l_degree}]{\includegraphics[scale=0.66]{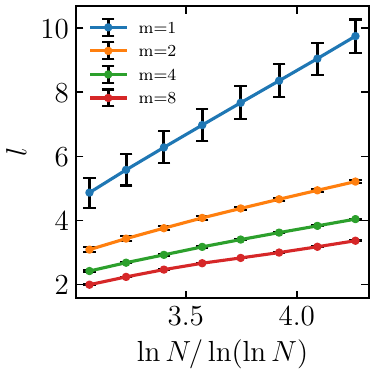}}\,
  \subfigure[\label{l_eigen}]{\includegraphics[scale=0.66]{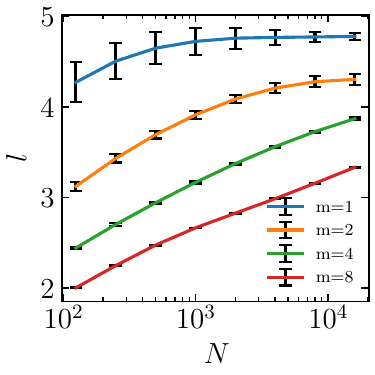}}
      \caption{Average shortest path length in terms of network size for eight values of $N$ including 125, 250, 500, 1000, 2000, 4000, 8000, and 16000, from left to right, with different values of $m$ for (a) BA model and (b) our model.}
    \label{l}
\end{figure}

The closeness of a node $i$ is defined as the reciprocal of its \textit{farness} relative to the other nodes multiplied by $N-1$,
\begin{equation}
C_i=\frac{N-1}{\sum_jd_{ij}},
\label{cls_eqt}
\end{equation}
so that
\begin{equation}
l=\frac{1}{N}\sum_i\frac{1}{C_i}.
\label{Eq:aspl-c}
\end{equation}
A node with higher closeness is more central in the sense that it is more accessible by the others, like super-hub in dandelion networks.
The average shortest path of the hub-and-spoke networks follows the statistics of star graph, which is trivially constant for large enough networks. More precisely, for a star graph one expects that $C_{\text{SH}}=1$ and $C_{ \text{nSH}}=\frac{N-1}{2N-3}$ (SH/nSH stands for super-hub/non-super-hub), so that $l=2\frac{N-1}{N}$, which goes to $C_{\text{SH}}=1$, $C_{ \text{nSH}}=\frac{1}{2}$ and $l=2$ as $N\to \infty$ (see SEC.~\ref{stargraph} for details). Therefore, based on the findings of the previous sections we expect that the average shortest path for the dandelion network saturates to a constant as the network size becomes sufficiently large. Figure~\ref{l} shows the average shortest path length as a function of the system size for BA and dandelion networks.  Fig.~\ref{l_degree} confirms the expected behavior for the BA model~\cite{fronczak2004average}, i.e. $l$ is linear in terms of $\frac{\ln N}{\ln(\ln N)}$. In the dandelion network for $m=1$, the average shortest path length tends to a constant, which is $l\to 4.78\pm 0.04$, showing that in spite of similarities, it is topologically different from star graph. This observation can be explained by the fact that the majority of newly added nodes in our model tend to connect to the super-hub node, so that the average shortest path length of the network will not surpass a certain value providing the network is sufficiently large. For $m>1$ it turns out that the average shortest path length saturates to some $m$-dependent values ($l\to 4.30\pm 0.03$ for $m=2$), while for a better characterization larger $N$ values are needed. The greater the value of $m$ in a network, the shorter its asymptotic average shortest path length.

\begin{figure*}[t]
  \centering
  \subfigure[\label{fig:viscloseness}]{\includegraphics[scale=0.3]{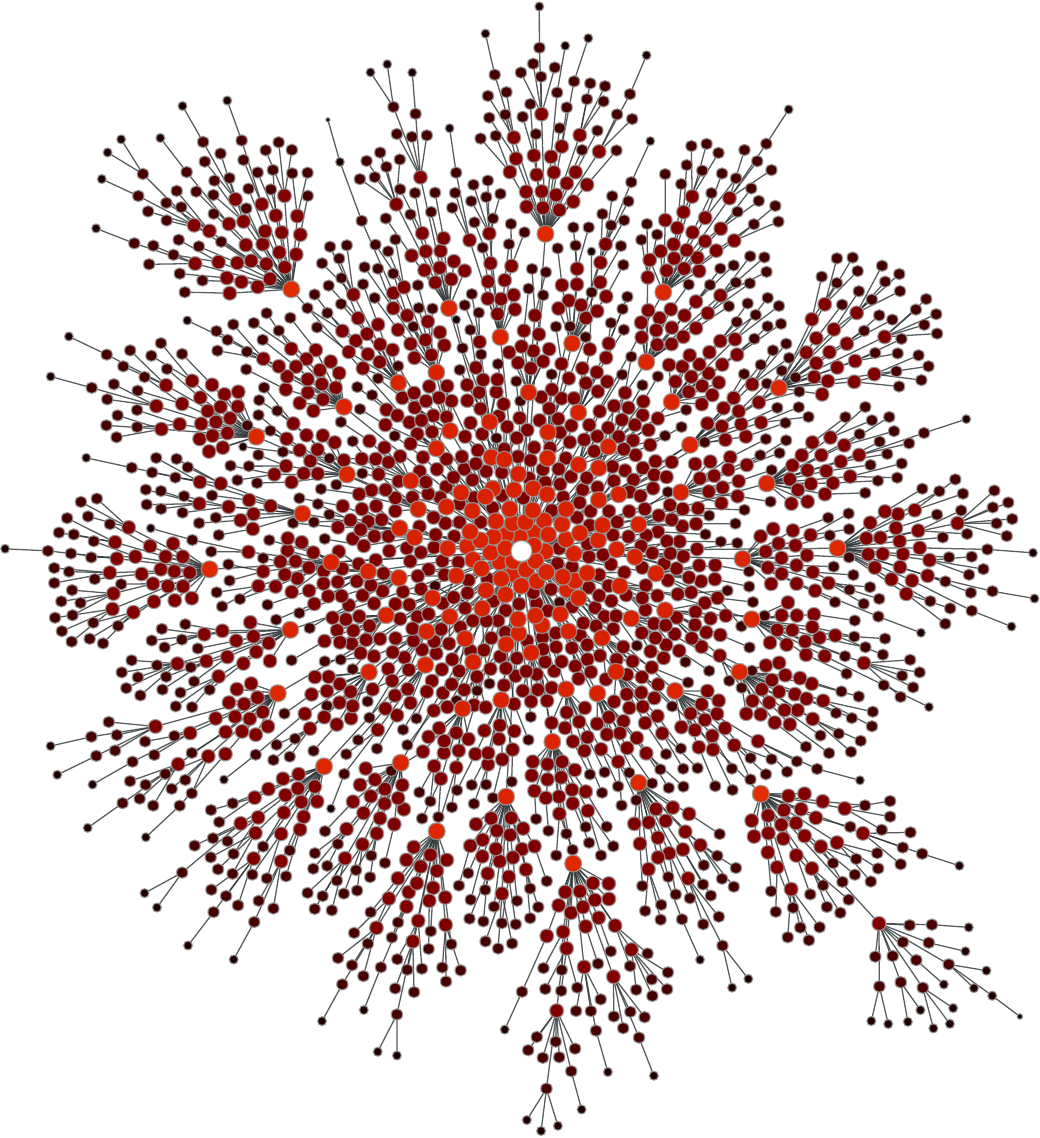}}\,
  \subfigure[\label{fig:cls_dist_ours}]{\includegraphics[scale=0.85]{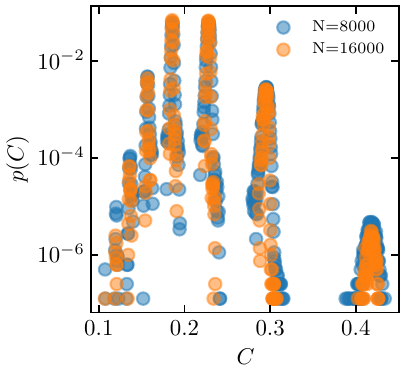}}\,
  \subfigure[\label{fig:cls_dist_BA}]{\includegraphics[scale=0.85]{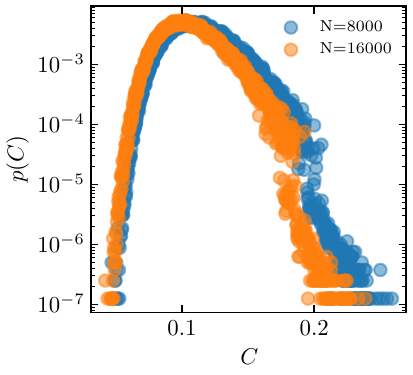}}
      \caption{(a) A representation of a network generated by our model with $N=2000$ and $m=1$. This visualization is based on different values of the nodes' closeness feature. Different sizes and colors are used to identify the lowest and highest values of closeness. The nodes with the smallest size and darkest color have the lowest value of closeness, whereas the brightest and largest nodes have the highest degree of closeness. The network's super-hub has the highest value of closeness. Closeness distribution of (b) our model and (c) BA model for different system sizes given $m=1$. While the BA model indicates a nearly continuous distribution, our model reveals a discrete one with some peaks at specified closeness amounts.}
    \label{fig:closeness}
\end{figure*}

Figure \ref{fig:viscloseness} depicts a $m=1$ dandelion network based on the nodes' closeness centrality with the size of points proportional to their closeness centrality. The network's super-hub has been identified by largest size and white color showing its highest closeness. A very different behavior between dandelion and BA for $m=1$ networks is evident in Figs.~\ref{fig:cls_dist_ours} and \ref{fig:cls_dist_BA}, the former showing a hierarchy based on their closeness centralities. The communities arisen from this closeness hierarchy is actually related to the hierarchical communities found based on the deletion of nodes, discussed in Sec.~\ref{SEC:community}. These peaks correspond to the branches of the dandelion; the $n$th peak corresponds to a $n$th level branch with distance $n$ form the super hub. Consider a node with distance $n$ from super-hub. The closeness centrality of the node $j$ distinct from the super-hub is given by
\begin{equation}
C_j^{(n)}\approx \frac{N-1}{\sum_k\left(d_{\text{SH}}^{(k)}+n\right)}=\frac{C_{\text{SH}}}{1+nC_{\text{SH}}},
\label{Eq:CSM-n}
\end{equation}
where $d_{\text{SH}}^{(k)}$ is the distance of node $k$ from the super-hub, and $C_{\text{SH}}=\frac{N-1}{\sum_jd_{\text{SH}}^{(j)}}$ is the closeness centrality of the super-hub. On the other hand the Eq.~\ref{Eq:aspl-c} tells us that
\begin{equation}
\begin{split}
Nl&=\frac{1}{C_{\text{SH}}}+\frac{n_1}{C_1}+\frac{n_2}{C_2}+...\\
&=\frac{1}{C_{\text{SH}}}\sum_{m=0}^Mn_m\left(1+mC_{\text{SH}}\right),
\end{split}
\end{equation}
where $n_m$ and $C_m$ and the number of nodes and the centrality of the nodes in the $m$th level, and $M$ is the maximum $m$ value available in a dandelion network with size $N$ (the total number of the communities). Noting that $\sum_{m=0}^Mn_m=N$, and that $n_m\propto N^{\frac{3}{4}m}$ (see Fig.~\ref{comm3}b) we obtain
\begin{equation}
\begin{split}
& NlC_{\text{SH}}=N\left(1+C_{\text{SH}}f(M)\right)\\
&\to C_{\text{SH}}=\frac{1}{l-f_N(M)},
\end{split}
\end{equation}
where
\begin{equation}
\begin{split}
f_N(M) & \equiv \frac{\sum_{m=0}^MmN^{\frac{3}{4}m}}{\sum_{m=0}^MN^{\frac{3}{4}m}}\\
&=\frac{MN^{\frac{3}{4}(M+2)}-(M+1)N^{\frac{3}{4}(M+1)}+N^{\frac{3}{4}}}{N^{\frac{3}{4}}(N^{\frac{3}{4}}-1)(N^{\frac{3}{4}M}-1)}.
\end{split}
\end{equation}
Noting that $\lim_{N\to\infty}f_N(M)\to M$, we finally find
\begin{equation}
C_{\text{SH}}=\frac{1}{l-M}, \ M<l.
\end{equation}
This equation, when combined with Eq.~\ref{Eq:CSM-n} gives ($M<l$)
\begin{equation}
C_{\text{SH}}^{(n)}=\frac{1}{l-M+n}, \ n=0,1,2,...,M.
\end{equation}
This turns out to be the discrete spectrum (the peak points) of the closeness centrality. For $l<M$ this argument fails since short connections prevent Eq.~\ref{Eq:CSM-n} to be valid.

\subsection{Re-scaling self-similarity}

\begin{figure}[h]
  \centering
  \subfigure[]{\includegraphics[scale=0.66]{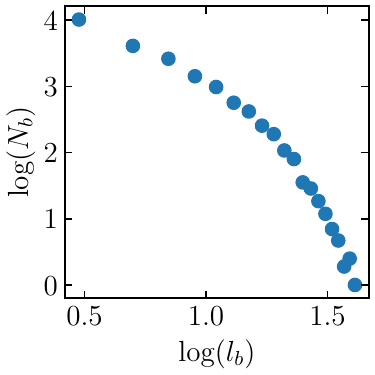}}\,
  \subfigure[]{\includegraphics[scale=0.66]{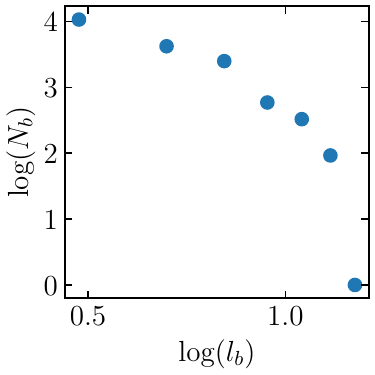}}
    \caption{None of the (a) BA model and (b) our model are self-similar. Here, we have used the random sequential algorithm ~\cite{kovacs2021comparative}. Evaluations are the result of 10 realizations for each model given $N=16000$ and $m=1$.}
    \label{fractal}
\end{figure}

The fractal nature of shapes is often assessed by using the ideas of fractality, fractal dimensions, and the box-counting technique. Fractal networks are based on the rules of fractal geometry, and their fractal dimension is calculated in a similar way as for shapes. The box-covering technique, also called box-counting, uses the shortest path between the points of a graph. This method works as follows for networks for a given characteristic \textit{box length} $l_b$:

\begin{itemize}
    \item A set of nodes is said to fit within a box of size $l_b$ if the shortest distance between any two of them is less than $l_b$. 
    \item The network is covered by a bunch of boxes of size $l_b$ if the nodes are divided so that each group fits into them. 
    \item The minimum number of boxes required to cover the network is identified as $N_b (l_b)$. 
    \item If the minimal number of boxes scales as a power of the box size, i.e., 
    \begin{equation}
        N_b (l_b)\propto l_b^{-d_b},
        \label{Eq:fractal}
    \end{equation}
    then the network is fractal with a finite fractal dimension or box dimension $d_b$.
\end{itemize}

We examine the fractal properties of both the BA and dandelion networks in this section. We use the random sequential algorithm \cite{kim2007box, kovacs2021comparative} to calculate the fractal dimension of the networks, among other methods. The random sequential algorithm uses the burning idea (breadth-first search). The boxes are created by burning (expanding) them from a randomly chosen center (or seed) node to its adjacent nodes. Moreover, nodes are "burned out" once they are allocated to a box.
At each level, a random unburned center node $s$ is selected, and then a sphere of radius $r_b$ (this radius is connected to $l_b$ as $l_b =2r_b+1$) is built around $s$; more specifically, the algorithm picks those unburned nodes that are no more than $r_b$ away from the center node $s$. These freshly burned nodes make up a new box.\\

The outcomes for the BA and dandelion networks are displayed in Figure \ref{fractal}. In a log-log plot, Eq.~\ref{Eq:fractal} implies that a linear graph should be obtained, whose slope is the fractal dimension. No linear pattern is discerned at any scale, which leads us to infer that the networks we examined are not fractal.

\section{Degree dynamics, mean field theory}~\label{degree_dynamics}

\begin{figure*}[t]
  \centering
  \subfigure[\label{fig:sigmai_t}]{\includegraphics[scale=0.6]{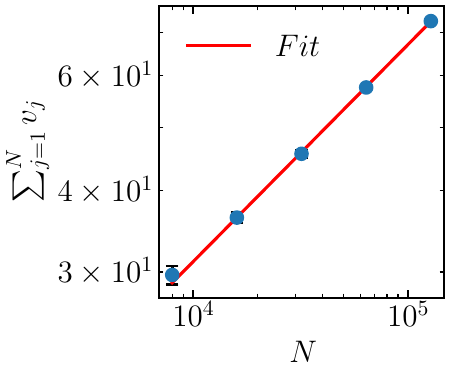}}\,
  \subfigure[\label{fig:kmax_2}]{\includegraphics[scale=0.6]{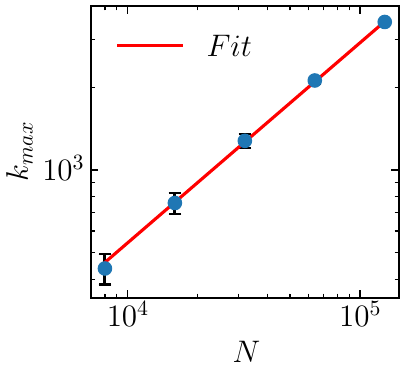}}\,
  \subfigure[\label{fig:vpeak_t}]{\includegraphics[scale=0.6]{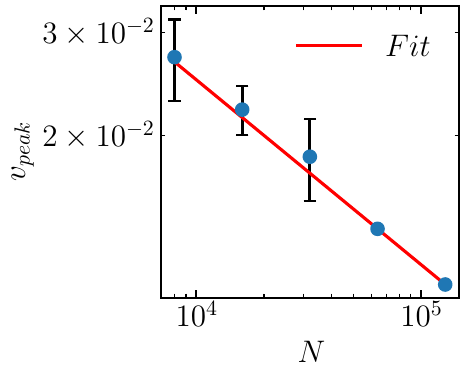}}\,
  \subfigure[\label{fig:degree_evo}]{\includegraphics[scale=0.6]{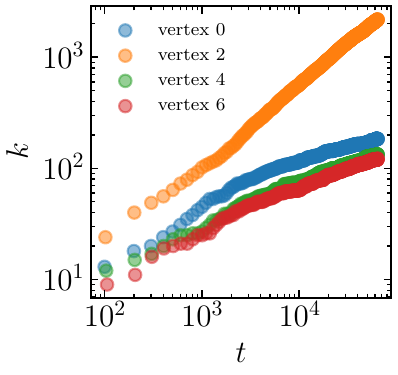}}
      \caption{(a) Variation of $\sum_{j=1}^N v_j$ as function of $N$ given $m=2$. There is a scaling realtion with the exponent $0.333 \pm 0.006$. (b) The growth of the hub's degree with time steps given $m=2$. The solid red line has the slope of $0.73\pm 0.03$. (c) The scaling relation of the $v_{peak}$ in terms of the size of the system given $m=2$. This scaling shown by solid red line reveals an exponent of $0.3159 \pm 0.0005$. (d) Degree evolution of some nodes a network with $N=64000$ and $m=2$. The nodes are added at time steps (indicating their indexes) 0, 2, 4 and 6. The node with index two (added at time step two) has had the chance to be chosen as the hub node. Here, one can easily spot the super-hub (node with index two) as its degree growth rate, $0.710\pm 0.001$, differs significantly from that of the other three nodes with an  average exponent of $0.296\pm 0.001$.} 
    \label{fig:sigmaj_kmax_vpeak_t}
\end{figure*}

In this section we focus on the dynamical aspects of the nodes on a mean field level. The average value of the degree $k_i$ is represented by a continuous real variable, the growth rate of which is expressed according to Eq.~\ref{Eq:ourpref}

\begin{equation}
	\frac{dk_i}{dt}\propto \frac{v_i}{\sum_j v_j}.
	\label{deg_evo_eqt}
\end{equation}

First we estimate $\sum_j v_j$ using the simulation results. Figure~\ref{fig:sigmai_t}, shows that there is a power-law relation between $\sum_j v_j$ and $N$ as

\begin{equation}
\sum_j v_j \sim N^{\alpha},
\label{N_32}
\end{equation}
with $\alpha = 0.333\pm 0.006$. This exponent, $\alpha \simeq \frac{1}{3}$, is $m$-independent and will play a crucial role in the network. Given that $v_{\text{SH}}\simeq \frac{\sqrt{2}}{2}$ one can solve Eq.\ref{deg_evo_eqt} as (knowing that $N\equiv t$)
\begin{equation}
k_{\text{max}}(t)\sim t^{\frac{2}{3}}
\label{Eq:kmax}
\end{equation}

This result in consistent with the numerical result for the time evolution of the super-hub's degree, as shown in figure \ref{fig:kmax_2}. For the nodes in the second level (distance 1 from the super-hub) the typical eigenvalue centrality (corresponding to the first peak point in the distribution function of the eigenvector centrality) runs with time approximately as (Fig.~\ref{fig:vpeak_t})
\begin{equation}
v_{\text{peak}}(t)\propto t^{-\frac{1}{3}}.
\end{equation}
We then are able to estimate the corresponding degree centrality ($k_{\text{\\peak}}$) as
\begin{equation}
	k_{\text{\\peak}}(t)\propto t^{\frac{1}{3}}.
\label{eqt_non_hub_degree_evo}
\end{equation}

This equation is valid for all non-super-hub nodes. We now understand the reason of the discontinuities seen in the figures \ref{fig:16000_ensemble} and \ref{fig:e_k_1}. In large enough networks (long enough time steps), the aforesaid discontinuities will increase as the degree of the super-hub grows faster than the degree of the other nodes. Figure \ref{fig:degree_evo} displays the rate of growth in the degree of the nodes inserted into the network at various time steps. One can easily spot the super-hub (node with index zero) here as its degree growth rate differs significantly from that of the other nodes. In the BA model, the rate at which the degree of the nodes changes is $\beta=0.5$ for all nodes but it is not the case here.

Note that, for $m>1$, more steps are required to see the discontinuities with respect to $m=1$. 

\begin{figure}[t]
  \centering
  \subfigure[]{\includegraphics[scale=0.61]{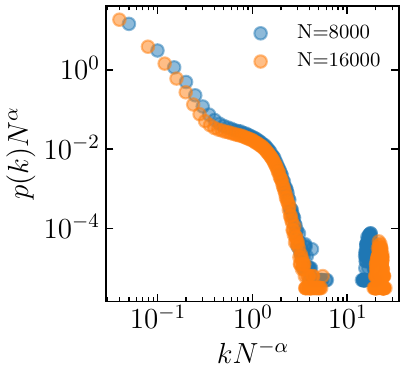}}\,
  \subfigure[]{\includegraphics[scale=0.61]{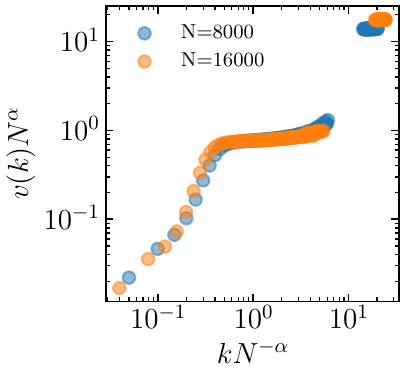}}
      \caption{At large enough networks, one can collapse the data representing the non-hub nodes in either (a) the degree distribution or (b) eigenvector-degree dependencies in one unite curve by plotting $f(k,N) N^{\alpha}$ against $k/ N^{\alpha}$ for different network sizes with $f(k,N)$ as degree distribution or eigenvector centrality of the nodes.}
    \label{collapsed}
\end{figure}

To confirm the results found in the mean field arguments, we use data collapse analysis. Figure \ref{collapsed} is analogous to the figures \ref{fig:16000_ensemble} and \ref{fig:e_k_1} in which we have scaled data reoresenting non-hub nodes with the scaling factor of $N^{-\alpha}$. We find the scaling relation as
\begin{equation}
f(k,N)\sim N^{-\alpha}\phi (k/ N^{\alpha})
\end{equation}
with $\alpha = 1/3$. $f(k,N)$ can represent either the probability distribution function of nodes' degree, $p(k,N)$, or eigenvector centrality of the nodes, $v(k,N)$. That is, if we plot $f(k,N) N^{\alpha}$ against $k/ N^{\alpha}$ for different network sizes, we will get a unite curve all data collapsed on it. The data representing the super-hubs will also collapse into a unite curve if we replace $N^{-\alpha}$ with $N^{\alpha-1}$ in the former equation.

\section{Concluding remarks}
In the Barab{\'a}si-Albert model, the nodes with high degree centralities have a good chance of being connected by new nodes, the paradigm known as the \textit{rich get richer}. There are however a lot of non-scale-free networks showing \textit{winner-takes-all} phenomenon, where only one super-hub appears in the network. The latter, called hub-and-spoke complex network, appears everywhere in various systems; air transportation networks, cargo delivery networks, telecommunication networks, and healthcare organization structures~\cite{toh1985impact, aykin1995networking, zapfel2002planning, klincewicz1998hub, elrod2017hub} are some instances of these networks. In this paper we propose an important class of preferential attachment complex networks showing winner-takes-all property based on the eigenvalue centrality which is reminiscent of a dandelion for $m=1$. One node becomes super-hub for long periods, which explains the opening of a gap in the degree centrality distribution.

As a non-scale-free network, our model displays unique characteristics such as a star-like pattern (where the CPD value approaches one in sufficiently large networks), a correlation between the degree of nodes and their clustering coefficient, and a distinct distribution of closeness unlike the continuous distribution seen in the BA model. Additionally, certain areas of degree centrality exhibit a hierarchical structure. Globally, the model maintains a steady average shortest path length over extended periods, qualifying it as a small-world network. We also evaluated the model for fractal characteristics and determined it does not possess this quality. Numerous statistical aspects of the model were analyzed, including betweenness, closeness, and h-index, from different statistical viewpoints; these are detailed in the tables \ref{tab:summary} and \ref{tab:corrcoef}. This model offers deeper understanding of the structure and behavior of real-world networks that display a hub-and-spoke structure.
\begin{table*}
\caption{\label{tab:summary}In this table we summarize the properties of the networks we found in our model and the BA model.}
\begin{ruledtabular}
\begin{tabular}{ccccccccc}
&scale-free&fractality&star graph&small-worldness&$l$&HS\footnote{hierarchical structure}&assortativity&CS\footnote{community structure}\\ \hline
Our model&in some regions&$\times$&\checkmark ($N\to \infty$)&\checkmark&const ($N\to \infty$)&in some regions&$\times$& $\times$ \\
BA model&\checkmark&$\times$&$\times$&\checkmark&$\sim\ln N/ \ln (\ln N)$&$\times$&$\times$&$\times$ \\
\end{tabular}
\end{ruledtabular}
\end{table*}

\begin{table*}
\caption{\label{tab:corrcoef}Correlation coefficient of different features in our model given $N=128000$ and $m=2$.}
\begin{ruledtabular}
\begin{tabular}{ccccccc}
&degree&h-index\footnote{In the academic context, the H-index~\cite{hirsch2005index} also known as the Hirsch index is determined as the highest number h for which there are h publications that have been cited at least h times each. On the other hand, within the realm of complex networks, a node's H-index is the highest number h where the node has h or more connections, each with a connection degree of at least h.}&eigenvector& betweenness&closeness&clustering coefficient\\ \hline
degree&$1.00$&-&-&-&-&- \\
h-index&$0.99$&$1.00$&-&-&-&- \\
eigenvector&$1.00$&$0.95$&$1.00$&-&-&- \\
betweenness&$1.00$&$0.96$&$1.00$&$1.00$&-&- \\
closeness&$0.97$&$0.96$&$0.77$&$0.99$&$1.00$&- \\
clustering coefficient&$-0.33$&$-0.84$&$-0.23$&$-0.85$&$0.05$&$1.00$ \\
\end{tabular}
\end{ruledtabular}
\end{table*}

\bibliography{refs}

\newpage

\appendix

\section{Degree distribution of the nodes of only one network generated by our model} ~\label{deg_dis_one}
As the network grows based on the eigenvector centrality of the nodes, one of the oldest nodes will draw most of the new incoming nodes to itself; the \textit{winner takes all} paradigm. Consequently, the degree of that node will rise with a much higher rate than the other nodes. Therefore, a discontinuity can be observed in the degree distribution of such networks. Figure \ref{fig:8000_sample} shows the degree distribution of a single sample of our model with size $N=16000$. The (degree of the) super-hub (\textit{the winner}) of the network can be easily recognized.

If we take another sample of the network, we will get qualitatively a similar degree distribution to that of figure \ref{fig:8000_sample}. If we take an average over enough number of realizations, we will get a degree distribution like the one in figure \ref{fig:16000_ensemble}, which can be said to have two parts.

It should be noted that the right part (with high degrees) in figure \ref{fig:16000_ensemble} is the average degree distribution of the super-hubs of each sample, while the left part is the degree distribution of the other nodes of those samples. It should also be noted that such clear discontinuities will only appear in large enough networks.

\section{Star graph} ~\label{stargraph}
\begin{figure*}[t]
  \centering
  \subfigure[\label{fig:8000_sample}]{\includegraphics[scale=0.65]{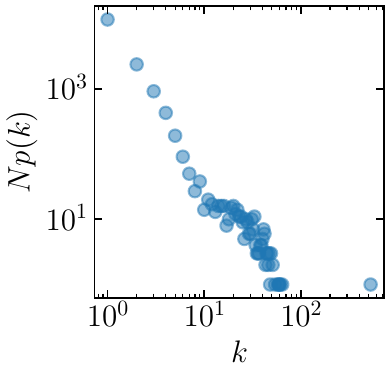}}\,  \subfigure[\label{fig:star_proof}]{\includegraphics[scale=0.084]{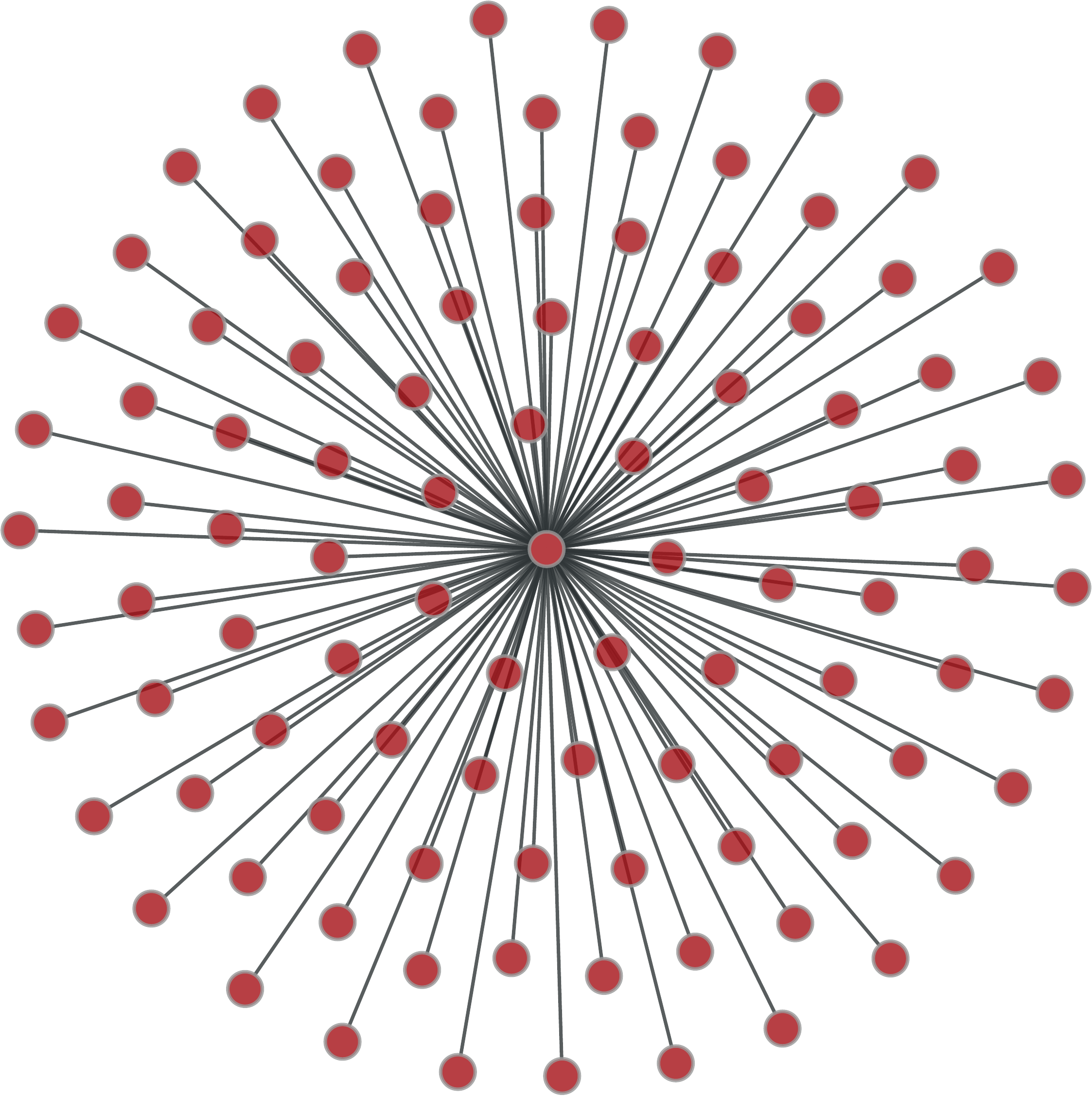}}\,
  \subfigure[]{\includegraphics[scale=0.14]{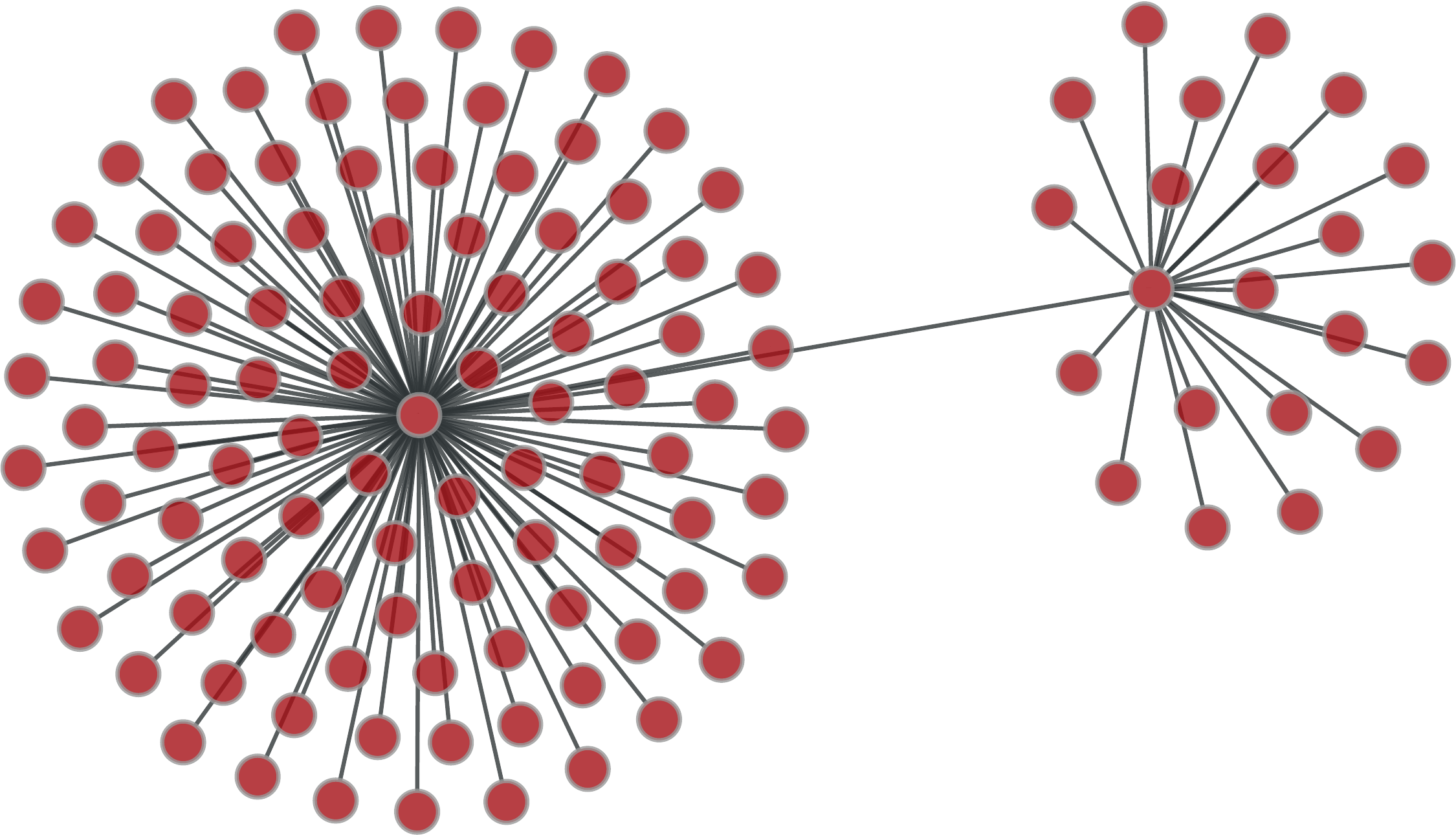}}\,  \subfigure[\label{fig:platto_proof}]{\includegraphics[scale=0.65]{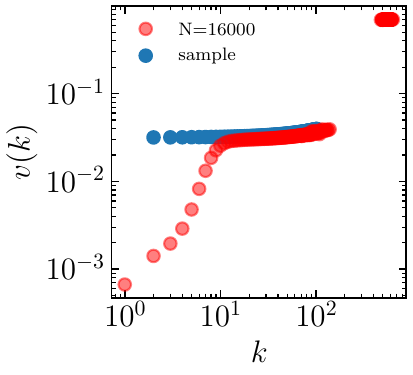}}
      \caption{(a) Degree distribution of a single sample produced by our model with parameters $N=16000$ and $m=1$. Notice the existence of a super-hub, i.e. the node with the highest degree in the system. (b) A star graph structure with a large hub at the center of the system. (c) Starting from the star graph in (b), we randomly pick one of the leaf nodes at each time step and connect new nodes to it. (d) We compute the eigenvector centrality of the nodes as we add new nodes to the selected node. We only show the eigenvector centrality of the selected node as a function of its degree by blue data points. By comparing the outcomes of this configuration (blue data points) with the outcomes of our model (red data points), one can infer that the flatness in the eigenvector centrality-degree relationship is due to the fact that the super-hub node is always the first neighbor of the nodes whose degrees form the flat region.}
    %\label{degree_dis}
\end{figure*}

A star graph of size $N=n+1$ ($n\geq 2$) is a tree graph that has one root and $n$ leaves; figure \ref{fig:star_proof}. The root has degree $n$ while each of the $n$ leaves has degree one. Therefore, the degree distribution of this structure has two bins, which means that this distribution is discrete.

The root node in a star graph has the highest values of degree (equal to $n$), eigenvector centrality (equal to $\frac{\sqrt{2}}{2}$), betweenness (equal to $1$), and closeness (equal to $1$) among the features. 

The CPD (central point dominance) is a feature of a star graph that always has a value of one, no matter the size of the network. The average shortest path length of a star graph depends on the network size; that is,
\begin{equation}
l=\frac{2n}{n+1}=\frac{2N-2}{N}.
\end{equation}
The value of $l$ approaches a maximum value of 2 as the network becomes very large.

\section{The flat region in the eigenvector-degree correlation} ~\label{flatness}
Figure \ref{fig:e_k_1} shows the eigenvector centrality of nodes as a function of their degrees. In networks that are large enough (with CPD values close to one), there is a range of degrees that have almost the same eigenvector centrality value as the degree of nodes changes in that range. That means, there is a flat region in the eigenvector centrality-degree phase space. We hypothesize that the nodes whose degrees are in that flat region have the largest node (super-hub) as their neighbours and this makes them have almost equal eignevector centrality.

\textbf{Proof}: First of all, we know (from our simulation results) that the eigenvector centrality of the super-hub is much higher than that of the other nodes. Second, In building our model we explained that the importance or eigenvector centrality of node $i$ is affected by the eigenvector centrality of its neighbours; mathematically
\begin{equation}
v_i=\frac{1}{\lambda_{max}}\sum_j a_{ij}v_j.
\label{flatness_proof}
\end{equation}

So, if node $i$ has the super-hub as its neighbour, one can simplify the above equation in networks that are large enough as
\begin{equation}
v_i \simeq \frac{1}{\lambda_{max}} a_{is}v_{s},
\end{equation}
with $s$ as the index of the super-hub. The eigenvector centrality of other neighbours of the node $i$ can be ignored compared to that of the super-hub. As a result, the eigenvector centrality of the nodes that have the super-hub as their neighbour will approach a constant number on average for networks that are large enough.

We test our conjecture by a configuration in which there is a large hub connected to nodes all of them having degree one (a star graph); figure \ref{fig:star_proof}. Then, by picking one of those leaf nodes that has the hub of the system as its neighbour, we start to link it to some new nodes. We keep doing this until the selected node's degree is almost $1/5$ (to compare with our simulation results) of the degree of the largest node. Figure \ref{fig:platto_proof} shows the eigenvector centrality-degree phase space in which the outcomes of the simulation based on our model and the outcomes of the recently described configuration are shown. The comparison of these two outcomes confirms our conjecture about the existence of the flat region. That is, the nodes whose degrees make up the flatness in the eigenvector-degree phase space have the super-hub as their neighbours on average. The starting point of the flatness corresponds to the average degree of the eigenvector centrality for which we see a peak in the eigenvector centrality distribution; the inset of the figure \ref{fig:ev_distribution_1}.

\end{document}